\documentclass[onecolumn]{article}
\usepackage{geometry}
\usepackage{graphicx} %use graph format
\usepackage{amssymb}  %Corresponding to 'draft' is 'final',which cancels the black line on the right
\usepackage{CJK}
\usepackage{amsmath}
\usepackage{amsthm}
\usepackage{amsfonts}
\usepackage{setspace}%Use spacing macro packages
\usepackage{tikz}%Draw macro package
\usepackage{bm}
\usepackage{dsfont}
\usepackage{graphicx}
\usepackage{subfigure}
\usepackage{cite}
\usepackage{booktabs}
\usepackage[hidelinks]{hyperref}

\title{
Ultrasonic Medical Tissue Imaging Using Probabilistic Inversion: Leveraging Variational Inference for Speed Reconstruction and Uncertainty Quantification
}
\author
{Qiang Li${}^{1}$, Heyu Ma${}^{1}$, Chengcheng Liu${}^{1,2,\ast}$, Dean Ta${}^{1,2,\ast}$\\
	\normalsize{${}^{1}$College of Biomedical Engineering, Fudan University, Shanghai 200438, China}\\
	\normalsize{${}^{2}$State Key Laboratory of Integrated Chips and Systems, Fudan University,}\\
	\normalsize{Shanghai 200438, China}\\
	\normalsize{$^\ast$Corresponding author. E-mail: chengchengliu@fudan.edu.cn;  tda@fudan.edu.cn}
}

\date{}

\begin{document}
\maketitle

\begin{abstract}
\textit{Background and Objective:}
Full Waveform Inversion (FWI) is a promising technique for achieving high-resolution imaging in medical ultrasound. 
However, conventional FWI methods suffer from issues related to computational efficiency, dependence on initial models, and the inability to quantify uncertainty.
This study aims to enhance inversion performance and provide a reliable method for uncertainty quantification in medical FWI imaging.

\textit{Methods:}
This study integrates the Stein Variational Gradient Descent (SVGD) algorithm into the FWI framework by deriving the posterior gradient for probabilistic inversion. 
To evaluate the proposed method, numerical experiments are conducted on synthetic datasets, including a breast tissue model with realistic anatomical structure.
Imaging accuracy and uncertainty quantification are assessed to compare the performance of SVGD-based FWI with conventional FWI and Stochastic Variational Inference (SVI) methods. Markov Chain Monte Carlo (MCMC) is implemented as a benchmark to evaluate the quality of uncertainty estimates.

\textit{Results:}
For synthetic data, the SVGD-based FWI framework yields more precise estimates in the region of interest (ROI) and demonstrates faster convergence compared to the conventional FWI.
For the anatomically realistic breast tissue simulation, SVGD produces a maximum relative error of 1.10\% and a mean relative error of 0.09\% in the ROI. The estimated uncertainty is spatially consistent, with most values below 0.01 and a mean of approximately 0.003. 
Compared to SVI, SVGD provides improved structural resolution and stronger agreement with the MCMC benchmark, indicating more reliable uncertainty quantification.

\textit{Conclusions:}
The SVGD-based FWI method improves inversion quality, enhances uncertainty quantification.
These findings indicate that probabilistic inversion is a promising tool for addressing the limitations of traditional FWI methods in ultrasonic imaging of medical tissues.

\end{abstract}

%\tableofcontents

\section{Introduction}
Ultrasound imaging plays an important role in diagnostics and monitoring in medical applications, providing real-time, non-invasive images of the internal body structures \cite{miao2023art,kim2015diagnostic,paoletta2021ultrasound,gao2024ultrasound}.
Despite its widespread applications, traditional B-mode ultrasound imaging has limited spatial resolution, making it difficult to clearly display complex tissue structures and heterogeneous media \cite{shrimali2009current,meiburger2018automated}.
Moreover, the B-mode imaging is highly susceptible to noise and artifacts, further compromising its diagnostic accuracy and robustness. 

In recent years, Full Waveform Inversion (FWI) has emerged as an advanced technique to enhance ultrasound imaging quality by utilizing the full ultrasonic waveform, rather than relying solely on amplitude or time-of-flight data \cite{feigin2020high, robins2021deep, suzuki2021optimized}. 
FWI iteratively adjusts model parameters to minimize the discrepancy between observed and simulated data, achieving high-precision estimations of medium properties \cite{lee2020seismic}. 
FWI has been proven to be both feasible and effective for inversing the speed of sound (SOS) image of musculoskeletal tissue \cite{li2021high, suo2021application,suo2023application} and brain tissues \cite{guasch2020full}, addressing challenges such as image distortion and low signal-to-noise ratio.
In quantitative imaging of bone tissues, FWI is capable of accurately reconstructing the fine structure of bone joints and capturing the cortical bone thickness \cite{bernard2017ultrasonic}. 

Traditional FWI algorithms face several challenges, including low computational efficiency, strong dependence on initial models, and difficulties in accurately quantifying parameter uncertainty \cite{yao2020review}. In uncertainty estimation for FWI, Markov Chain Monte Carlo (MCMC) methods, such as the Metropolis-Hastings algorithm \cite{chib1995understanding}, have traditionally been used to sample from complex posterior distributions \cite{geyer1992practical,brooks1998markov}. 
However, MCMC methods are inefficient in high-dimensional spaces due to their random-walk behavior, which leads to slow convergence and the tendency to become trapped at individual maxima in the probability distribution \cite{Zhang_Curtis_2024,vrugt2011dream}. 
Although more advanced MCMC techniques, such as reversible-jump MCMC \cite{green2009reversible,waagepetersen2001tutorial,zhang20183}, stochastic Newton MCMC\cite{martin2012stochastic,petra2014computational}, and Hamiltonian Monte Carlo \cite{chen2014stochastic,betancourt2017conceptual}, aim to improve efficiency, 
these methods remain intractable for FWI due to their extremely high computational cost.

To improve the efficiency of MCMC methods, Variational Inference (VI) techniques are increasingly applied to estimate the posterior distribution of model parameters and address uncertainties inherent in the inversion process \cite{zhao2024physically,yin2024wise,zhang2020variational}. Instead of directly calculating the posterior distribution, VI approximates it by minimizing the Kullback-Leibler (KL) divergence between the true posterior and a simpler, tractable distribution \cite{kullback1951information,zhang2018advances}. Recent research, such as stochastic variational inference (SVI) for medical ultrasound imaging \cite{bates2022probabilistic}, has demonstrated how gradient-based methods can efficiently solve inverse problems. The derivation of SVI depends on selecting a suitable prior distribution, which can be challenging to determine in practice. Stein Variational Gradient Descent (SVGD), a VI-based method, uses a set of particles to approximate more complex prior distributions, such as a Gaussian mixture model \cite{liu2016stein}. The SVGD approach enhances both the efficiency and accuracy of VI in high-dimensional inverse problems, making it a promising method for addressing challenges in ultrasound imaging.

This paper explores the application of SVGD in medical ultrasound imaging, focusing on enhancing the efficiency and robustness of the inversion process.
The core principles of the wave equation used in FWI to model wave propagation are outlined, followed by a discussion of the challenges in model inversion due to noise and complex data.
The paper then introduces VI as a method for incorporating uncertainty into the inversion process, improving upon traditional optimization techniques.
SVGD is presented as an advanced VI-based algorithm that optimizes the approximation of posterior distributions.
The gradient of the posterior distribution is derived, and SVGD is applied to solve inverse problems in medical ultrasound imaging. 
Simulation results on both synthetic phantoms and anatomically realistic breast models demonstrate that SVGD improves convergence speed, reconstruction accuracy, and the reliability of uncertainty estimates.

The structure of this paper is outlined below.
Section \ref{sec: methods} introduces the foundational concepts of FWI and VI, with a focus on the application of SVGD to FWI, including the derivation of the posterior gradient.
In Section \ref{sec: results}, the proposed framework is applied to synthetic data generated from both linear-array and ring-array transducers, and further extended to breast tissue data.
This section uses conventional FWI as a baseline to evaluate the inversion quality of SVGD, and employs SVI and MCMC for a comparative analysis of the uncertainty quantification provided by SVGD.
Finally, Sections \ref{sec: discussion} and \ref{sec: conclusion} provide a discussion of the results and conclude the study, respectively.

\section{Methods} \label{sec: methods}
\subsection{Acoustic wave propagation and data generation in FWI}
The wave equation in FWI is used to simulate wave propagation and iteratively update the model to match the observed data. 
The most commonly used wave equation is the acoustic wave equation \cite{pratt1990inverse,pierce2019acoustics},
\begin{equation} \label{wave eq src}
	\frac{1}{\bm c^2}\frac{\partial^2 \bm p(\bm{x}, t)}{\partial t^2} - \nabla^2 \bm p(\bm{x}, t) = \bm s(\bm{x}, t),
\end{equation}
where $\bm p(\bm{x}, t)$ represents acoustic pressure at position $\bm{x} = (x_1, x_2, \cdots, x_d)$ and time $t$. 
The notation $\bm c= \bm c(\bm x)$ is the speed of sound (SOS), which varies spatially in a heterogeneous medium.
In FWI applications, the model parameter is often reparameterized as $\bm m = 1/\bm c^2$, known as the squared slowness, to facilitate gradient-based optimization and improve numerical convergence.
The operator $\nabla^2 = \sum_i \partial^2/ \partial x_i^2 $ is the Laplacian operator, representing the spatial derivatives of the pressure field. 
The source signal $\bm  s(\bm{x}, t)$ describes the spatial and temporal distribution of the energy input into the system. 
The exact form of $\bm s(\bm{x}, t)$ depends on the specific modeling of the ultrasound source. For a simple point source, the expression is
\begin{equation*}
	\bm s(\bm{x}, t) = \delta(\bm{x}-\bm{x}_s) \bm f(t),
\end{equation*}
where $\delta(\bm{x}-\bm{x}_s)$ is the Dirac delta function, representing a point source located at position $\bm{x}_s$. The function $\bm f(t)$ is a time-dependent function that models the temporal profile of the source, such as a pulse or continuous wave.

The receivers, positioned at $\bm{x}_r$, convert the mechanical vibrations of the incoming acoustic waves into electrical signals. 
These signals consist of both reflected and transmitted waves. 
To predict the wavefield data recorded by the receivers, the wave equation (\ref{wave eq src}) is numerically solved using the finite difference method.
The obtained data $\bm{d}_{r}$ is a time series and described by a functional relation that involves the position $\bm{x}_r$, time $t$, and squared slowness $\bm{m}$. This relationship is
\begin{equation} \label{nonlin map wave}
	\bm d_{r} = \mathcal{L}(\bm x_r, t, \bm m),
\end{equation}
where $\mathcal{L}$ is a nonlinear operator that models the conversion of the mechanical displacement (or wavefield) into electrical signals by the transducer.

While the predicted data $\bm{d}_r$ is obtained by solving the wave equation with the parameters $\bm{m}$, real-world measurements $\bm{d}_r^{\text{obs}}$ typically exhibit discrepancies when compared to $\bm{d}_r$. 
The discrepancies are modeled as
\begin{equation} \label{nonlin map error}
	\bm{d}_r^{\text{obs}} = \mathcal{L}(\bm{x}_r, t, \bm{m}) + \bm{\epsilon},
\end{equation}
where $\bm{\epsilon}$ captures errors arising from an un-converged SOS model, as well as factors including sensor imperfections, environmental conditions, and inherent measurement noise. 
Therefore, the observed data $\bm{d}_r^{\text{obs}}$ is often a noisy and potentially distorted version of the predicted data $\bm{d}_r$, and addressing these discrepancies is essential for accurate model inversion and updating in FWI.

FWI is a powerful computational technique used to create high-resolution models of tissue properties. 
Unlike traditional B-mode methods that rely on simplified approximations, FWI makes use of the full information contained in the entire waveform of the received signal, including both amplitude and phase information. 
The primary objective of FWI is to iteratively update the model parameters $\bm m$  by comparing the experimentally observed data $\bm{d}_r^{\text{obs}}$ with the predicted data $\bm{d}_r$. 

FWI begins with an initial model of the medium, which serves as an approximation of the model parameters $\bm{m}$ at each spatial point. Using this initial model, the predicted wavefield data $\bm{d}_r$ is computed numerically by solving the wave equation (\ref{wave eq src}). 
The core of FWI is the iterative updating of the parameters $\bm{m}$ by minimizing the difference between $\bm{d}_r^{\text{obs}}$ and $\bm{d}_r$.
The difference between the two datasets is quantified using a misfit function, often based on the least-squares error. 
The most commonly used misfit function is the $l_2$ norm, which measures the sum of squared differences across all receivers \cite{bates2022probabilistic,wu2024full,wu2023ultrasound},
\begin{equation} \label{l2 misfit}
	l_2 (\bm{d}_r^{\text{obs}}, \bm{d}_r) = \frac{1}{2} \sum_r \|\bm{d}_r-\bm{d}_r^{\text{obs}}\|^2.
\end{equation}
The goal of FWI is to find the optimal parameters $\bm{m}^*$ that minimize the misfit function, i.e., to solve the following optimization problem,
\begin{equation} \label{min l2}
	\bm m^* = \mathop{\arg\min}_{\bm m}\ l_2 (\bm{d}_r^{\text{obs}}, \bm{d}_r).
\end{equation}
In practice, this minimization is achieved through iterative optimization techniques. The optimization process is typically nonlinear, as the relationship between the model parameters and the observed data is often complex and involves multiple reflections, scattering, and other high-order wave phenomena.

One of the key challenges of FWI is the possibility for the optimization to converge to local minima, especially when the initial model is far from the true solution. To mitigate this issue, careful initialization and regularization techniques are often employed. Additionally, FWI is computationally expensive, as both the forward and adjoint problems need to be solved repeatedly during the inversion process. Despite these challenges, FWI provides a powerful framework for obtaining high-resolution, accurate models of the medium's properties, which significantly improve imaging and characterization in applications.

\subsection{Variational inference solves FWI problems}

In recent years, VI has emerged as a powerful tool for improving the efficiency and robustness of inversion techniques. VI provides an alternative approach to classical optimization by approximating complex posterior distributions with simpler, tractable distributions, thus reducing the computational burden. The goal of FWI is to find the optimal parameters $\bm{m}^*$ satisfying Eq. (\ref{min l2}), VI treats the parameters $\bm{m}$ as random variables and infers their distribution given the observed data. 
%This probabilistic framework allows the incorporation of uncertainty into the inversion process, offering not only an optimal estimate of the model parameters but also a measure of the confidence in these estimates.
This allows for more reliable reconstructions of tissue properties, providing both the best estimate of model parameters and a measure of uncertainty that is crucial for clinical decision-making.
By approximating the true posterior distribution through a variational family, VI facilitates efficient model updating and allows for more robust handling of noisy or incomplete data. 
This is beneficial in medical ultrasound imaging, where measurement noise, sensor limitations, and complex tissue structures often make traditional deterministic inversion methods prone to local minima or overfitting. VI-based methods can enhance the accuracy and stability of FWI, providing more reliable reconstructions of tissue properties.

Bayes' theorem provides the foundational framework for VI-based methods,
which update the parameters $\bm{m}$ based on observed data $\bm{d}^{\text{obs}}$. 
Bayes' theorem  expresses the posterior distribution $p(\bm{m} | \bm{d}^{\text{obs}})$ as
\begin{equation} \label{bayes them}
	p(\bm{m} | \bm{d}^{\text{obs}}) = \frac{p(\bm{d}^{\text{obs}} | \bm{m}) p(\bm{m})}{p(\bm{d}^{\text{obs}})},
\end{equation}
where $p(\bm{m})$ is the prior distribution that encodes our initial beliefs about the model parameters before observing the data.
The evidence distribution $p(\bm{d}^{\text{obs}})$ acts as a normalizing constant ensuring that the posterior distribution integrates to one. 
The likelihood distribution $p(\bm{d}^{\text{obs}} | \bm{m})$ represents the probability of observing the data given a set of model parameters.
Although Eq.(\ref{l2 misfit}) does not explicitly include a regularization term, such terms are incorporated in practice to mitigate ill-posedness and improve inversion stability. 
In the Bayesian framework, regularization naturally arises through the prior distribution, where the negative logarithm of the prior acts as a regularizer. 
This probabilistic interpretation provides a principled way to incorporate prior knowledge into the inversion.

A widely used numerical method to approximate the posterior distribution in Eq. (\ref{bayes them}) is MCMC \cite{chib1995understanding,jones2022markov}. 
MCMC is a class of algorithms that generates samples from a distribution proportional to the posterior, i.e.,
$p(\bm{m}|\bm{d}^{\text{obs}}) \propto p(\bm{d}^{\text{obs}}|\bm{m})p(\bm{m})$.
The samples allow for estimating expectations and quantifying uncertainty without explicitly computing the normalization constant.

Although MCMC provides an asymptotically correct approximation of the posterior, it can be computationally intensive and inefficient in high-dimensional inverse problems such as FWI. 
To address these limitations, VI approximates the posterior distribution $ p(\bm{m} | \bm{d}^{\text{obs}}) $ by minimizing the KL divergence between the true posterior and a simpler, tractable variational distribution $ q(\bm{m}) $. The variational distribution $ q(\bm{m}) $ belongs to a chosen family of distributions $ \mathcal{Q} $, and the optimal distribution $ q^*(\bm{m}) $ is obtained by 
\begin{equation} \label{q* KLdiv}
	q^*(\bm{m}) = \mathop{\arg\min}_{q(\bm m) \in \mathcal Q} D_{KL}\left[q(\bm m) \| p(\bm{m} | \bm{d}^{\text{obs}})\right],
\end{equation}
where the KL divergence is defined as
\begin{align} \label{KL div}
	D_{KL} \left[q(\bm m) \| p(\bm{m} | \bm{d}^{\text{obs}})\right] = & \int q(\bm m) \log \left(\frac{q(\bm m)}{p(\bm{m} | \bm{d}^{\text{obs}})}\right) d \bm{m} \nonumber  \\
	= & \int q(\bm m) \log \left(q(\bm m)\right)d \bm{m} - \int q(\bm m) \log \left(p(\bm{m} | \bm{d}^{\text{obs}})\right) d \bm{m}.
\end{align}
The KL divergence is non-negative and measures the difference between two probability distributions \cite{belov2011distributions,zhang2024properties}. 
%Although the proof of the non-negativity of the KL divergence is straightforward, we provide a proof in \textit{Appendix A} for completeness.

By leveraging the KL divergence, VI reduces the computational burden compared to exact Bayesian inference methods, such as MCMC methods. 
The flexibility of VI lies in the choice of optimization techniques and the variational family $\mathcal{Q}$, which leads to the development of various algorithms.
Among these, SVGD stands out as a flexible and efficient method.
In the following, this paper introduces SVGD and demonstrates its efficiency in solving inverse problems.

\subsection{SVGD-based posterior approximation and uncertainty quantification in FWI} 

SVGD is an advanced algorithm designed for approximating complex posterior distributions in VI\cite{liu2016stein}. 
The SVGD algorithm forms a natural counterpart to gradient descent for optimization problems. Instead of relying on a fixed set of particles, SVGD actively transports particles to better match the target distribution by leveraging functional gradient descent.

The core idea of SVGD is to apply an incremental transformation to the model parameters $ \bm{m} $. This transformation is defined by the identity map \cite{liu2016stein}
\begin{equation} \label{1to1 map}
	f(\bm m)  = \bm m + \epsilon \, \bm g(\bm m),
\end{equation}
where $ \bm{g}(\bm{m}) $ is a smooth function that represents the perturbation direction, and $ \epsilon $ is a scalar that controls the magnitude of the perturbation. This transformation is designed to be an invertible map, satisfying the inverse function theorem, and gradually updates the approximate distribution to match the true posterior. 

The density of the transformed distribution is denoted as $ q_f(z) $, where $ z = f(\bm{m}) $ is the transformed model.
To optimize the approximation, SVGD minimizes the KL divergence between the transformed distribution $ q_f(z) $ and the true posterior $ p(\bm{m} | \bm{d}^{\text{obs}}) $. The gradient of the KL divergence with respect to $ \epsilon $ is
\begin{equation} \label{grad D_KL=-Eq}
	\left.\nabla_\epsilon D_{KL}\left[q_f \| p\right]\right|_{\epsilon=0}=-E_q\left[\operatorname{trace}\left(\mathcal{A}_p \bm g(\bm{m})\right)\right],
\end{equation}
where $ \mathcal{A}_p $ is the Stein operator, expressed as
\begin{equation}
	\mathcal{A}_p \bm{g}(\bm{m})=\nabla_{\mathrm{m}} \log p\left(\bm{m} \mid \bm{d}_{\mathrm{obs}}\right) \bm{g}(\bm{m})^T+\nabla_{\mathrm{m}} \bm{g}(\bm{m}) .
\end{equation}
Eq. (\ref{grad D_KL=-Eq}) describes how the perturbation function $ \bm{g}(\bm{m}) $ is updated to reduce the KL divergence between the approximate distribution and the true posterior. The gradient of the KL divergence informs the direction to move the model parameters in order to improve the approximation.

To find the optimal perturbation direction, Ref. \cite{liu2016stein} derives the steepest descent that maximizes the negative gradient in Eq. (\ref{grad D_KL=-Eq}) as
\begin{equation}
	\bm g^*(\bm m)= E_{\bm m \sim q}\left[k(\bm m, \bm m') \nabla_{\bm m} \log p\left(\bm{m} \mid \bm{d}_{\mathrm{obs}}\right) +\nabla_{\bm m} k(\bm m, \bm m')\right],
\end{equation}
where $ k(\bm{m}, \bm{m}') $ is a kernel function that quantifies the similarity between model parameters $ \bm{m} $ and $ \bm{m}' $. 
The kernel function determines how the perturbation at one point influences nearby points in the model space. One common choice for the kernel function is the Radial Basis Function (RBF) kernel,
\begin{equation}
	k(\bm{m}, \bm{m}') = \exp\left( -\frac{\|\bm{m} - \bm{m}'\|^2}{2 h^2} \right),
\end{equation}
where $ h $ is a hyperparameter that controls the width of the kernel and determines the influence of one point on another. The RBF kernel is widely used due to its smoothness and locality, ensuring that points that are closer in model space will have more influence on each other. The gradient of the kernel function with respect to $ \bm{m} $ is
\begin{equation}
	\nabla_{\bm{m}} k(\bm{m}, \bm{m}') = -\frac{\bm{m} - \bm{m}'}{\sigma^2} \exp\left( -\frac{\|\bm{m} - \bm{m}'\|^2}{2 \sigma^2} \right).
\end{equation}
The final update rule at each iteration is
\begin{align} \label{svgd update rule}
	\bm{m}_j^{t+1} =\bm{m}_j^t+\frac{\epsilon^t}{n}\sum_{i=1}^{n} \left[k\left(\bm{m}_i^t, \bm{m}_j^t\right) \nabla_{\bm{m}_i^t} \log p\left(\bm{m}_i^t \mid \bm{d}^{\mathrm{obs}}\right)+\nabla_{\bm{m}_i^t} k\left(\bm{m}_i^t, \bm{m}_j^t\right)\right].
\end{align}
Where $n$ is the number of particles.
The above iterative update process, using the kernel function and the gradient of the log posterior, refines the approximation to the true posterior distribution. 

%\subsection{Gradient calculation of log-posterior for SVGD method}

To apply the SVGD update rule in FWI, by substituting Bayes' theorem (\ref{bayes them}) into the gradient of the log-posterior (\ref{svgd update rule}), yields
\begin{align}
	\nabla_{\bm{m}} \log p\left(\bm{m} \mid \bm{d}^{\mathrm{obs}}\right) = & \nabla_{\bm{m}}\log p\left(\bm{d}^{\mathrm{obs}} \mid \bm{m}\right)+\nabla_{\bm{m}}\log p\left(\bm{m}\right).
\end{align}
In the process of FWI, the likelihood term $ p(\bm{d}^{\text{obs}} | \bm{m}) $ is typically modeled using a misfit function, such as the least-squares error (\ref{l2 misfit}). Using this least-squares error formulation, the Gaussian likelihood function is expressed as \cite{bates2022probabilistic},
\begin{equation} \label{MLE=exp(misfit)}
	p\left(\bm{d}^{\mathrm{obs}} \mid \bm{m}\right) = \operatorname{trace}(2 \pi \mathbf I)^{-\frac{1}{2}} \exp \left\{-\frac{1}{2} \sum_r \|\bm{d}_r-\bm{d}_r^{\text{obs}}\|^2\right\}.
\end{equation}
where $ \mathbf{I} $ is the identity matrix.
The exponential term in Eq. (\ref{MLE=exp(misfit)}) cancels with the logarithm,
\begin{equation} \label{logMLE=misfit}
	\log p\left(\bm{d}^{\mathrm{obs}} \mid \bm{m}\right) \propto -\frac{1}{2} \sum_r \|\bm{d}_r-\bm{d}_r^{\text{obs}}\|^2=-l_2(\bm{d}_r^{\text{obs}}, \bm{d}_r) .
\end{equation}

To obtain $\nabla_{\bm{m}}\log p\left(\bm{d}^{\mathrm{obs}} \mid \bm{m}\right)$, 
the gradient of the misfit function $ l_2 $ with respect to $ \bm{m} $ is required.
The gradient is given by
\begin{equation}
	\nabla_{\bm{m}} l_2(\bm{d}_r^{\text{obs}}, \bm{d}_r) = \sum_r \left( \bm{d}_r - \bm{d}_r^{\text{obs}} \right) \cdot \nabla_{\bm{m}} \bm{d}_r.
\end{equation}
Since $ \bm{d}_r $ depends on the model parameters $ \bm{m} $, it is necessary to compute $ \nabla_{\bm{m}} \bm{d}_r $, which involves differentiating the forward model $ \mathcal{L} $ with respect to $ \bm{m} $.

To avoid direct computation of the gradient of $ \bm{d}_r $ with respect to $ \bm{m} $ (which is computationally challenging), the adjoint state method \cite{fichtner2006adjoint} is used.
The adjoint state, denoted as $ \bm{a}_r $, is an auxiliary variable introduced to propagate the error backward through the system. 
By solving the backward equation for the adjoint state, the gradient of the misfit function is computed without the need to differentiate the forward model

The adjoint state $ \bm{a}_r $ satisfies a backward equation, which is derived by applying the chain rule to the misfit function $l_2$. The adjoint equation is 
\begin{equation}
	\mathcal{L}^\dagger  \bm{a}_r = \bm{d}_r - \bm{d}_r^{\text{obs}},
\end{equation}
where $ \mathcal{L}^\dagger  $ is the adjoint of the forward operator $ \mathcal{L} $. The adjoint state $ \bm{a}_r $ propagates the error backward from the data space to the model space.
Once the adjoint state $ \bm{a}_r $ is obtained, the gradient of the misfit function with respect to the parameters $ \bm{m} $ is
\begin{equation} \label{l2dm adjoint}
	\nabla_{\bm{m}} l_2(\bm{d}_r^{\text{obs}}, \bm{d}_r) = \sum_r \bm{a}_r \cdot \nabla_{\bm{m}} \mathcal{L}(\bm{x}_r, t, \bm{m}).
\end{equation}

To evaluate the gradient of $\log p(\bm{m})$, the prior distribution $p(\bm{m})$ is defined as a truncated Gaussian with bounded support $[m_{\min}, m_{\max}]$, ensuring that $\bm{m}$ remains within its physically plausible range.
Following the mean-field parameterization strategy \cite{bates2022probabilistic,zhang20233}, the prior is expressed as
\begin{equation}
	\label{p(m)=prod p(mi)} p(\bm{m}) = \prod_{i=1}^d p(m_i), \quad m_i \in [m_{\min}, m_{\max}].
\end{equation}
Each marginal $p(m_i)$ is specified as a truncated Gaussian,
\begin{equation} \label{gaussian prior}
	p(m_i) \propto \frac{1}{\sqrt{2\pi}\sigma_i}
	\exp\!\left(-\frac{(m_i - \mu_i)^2}{2\sigma_i^2}\right),
\end{equation}
normalized by the cumulative probability within the truncation interval.
The mean and variance of the $ i $-th component $ m_i $ are denoted by 
$ \mu_i $ and $ \sigma_i^2 $, respectively. 
The factorization (\ref{p(m)=prod p(mi)}) reduces the complexity of the prior distribution by assuming independence across parameters.
The gradient of $ \log p(\bm{m}) $ with respect to each $ m_i $ is 
\begin{equation} \label{logp dm}
	\nabla_{m_i}\log p(\bm{m}) = -\frac{m_i - \mu_i}{\sigma_i^2}, 
	\quad m_i \in [m_{\min}, m_{\max}].
\end{equation}
Stacking the gradients for all components $ m_i $ results in  the gradient $ \nabla_{\bm{m}} \log p(\bm{m} \mid \bm{d}^{\mathrm{obs}})$.

\section{Results} \label{sec: results}

\begin{figure*}[ht]
	\centering
	\includegraphics[scale=0.4]{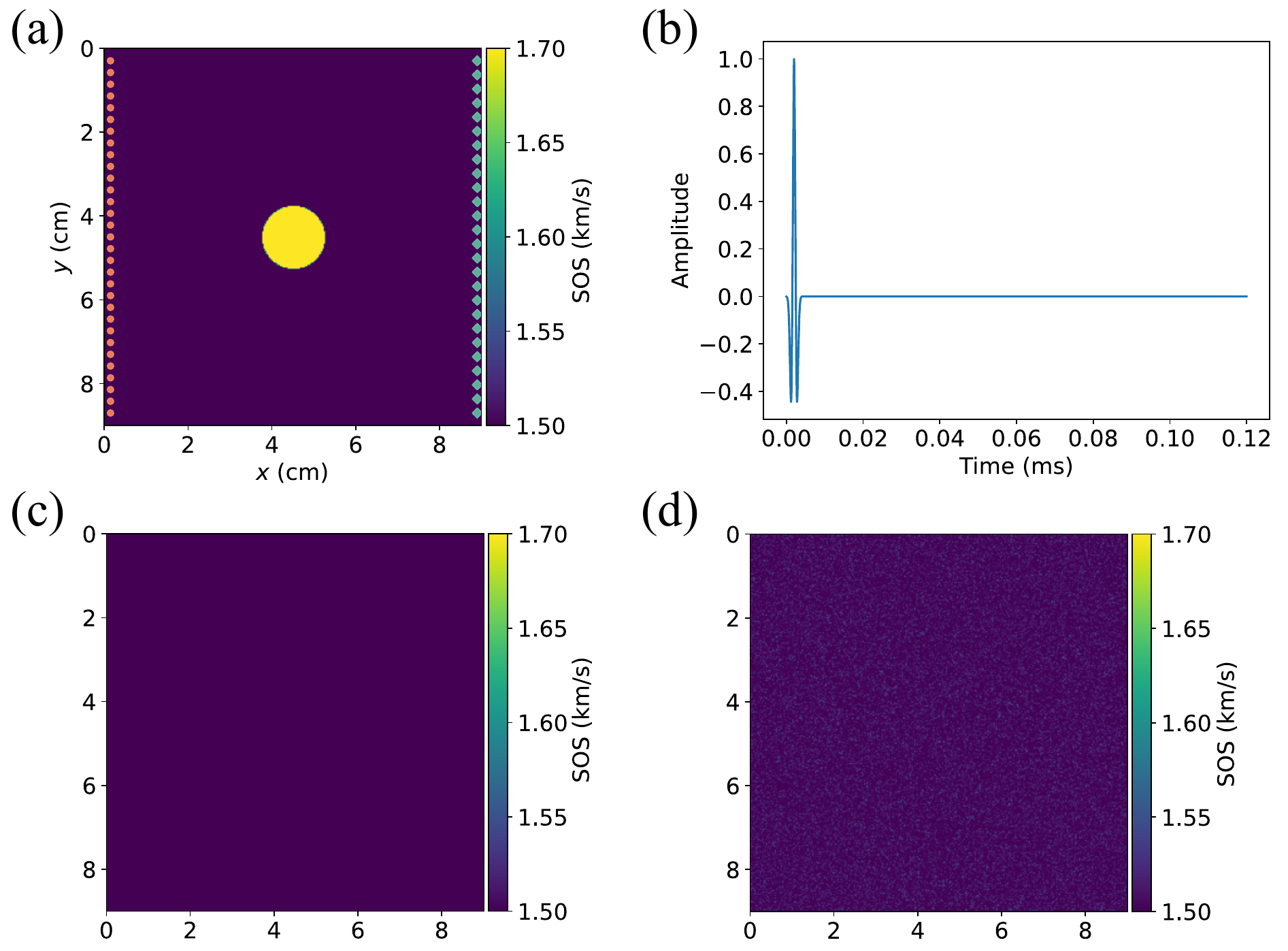}
	\caption{ Inversion setup and initial conditions for the synthetic data experiment using a linear array transducer. 
	(a) The true model setup. The orange dots and green diamonds indicate the locations of the sources and receivers, respectively.
	(b) The Ricker wavelet used as the source waveform.
	(c) The initial model for the conventional FWI.
	(d) An example of the initial particles for SVGD, with a total of 20 particles used in the inversion.
	}
	\label{fig:fwi2DlineSvgdInit}
\end{figure*}

In this section, the gradients in Eqs. (\ref{l2dm adjoint}) and (\ref{logp dm}) are combined to apply the SVGD update rule (\ref{svgd update rule}) for solving FWI problems under different experimental conditions.
The primary focus is on the SOS distribution of objects or tissues within the acoustic field. 
Using the obtained SOS information, imaging of the objects or tissues is performed. 
%To assess the convergence of the SVGD algorithm, Eq. (\ref{l2 misfit}) is used as the loss function. 
Once the update steps converge, the SOS distribution is visualized by plotting the mean of the particles, and the uncertainty in the corresponding SOS is measured through the standard deviation. 
Additionally, to evaluate the effectiveness of SVGD, the conventional FWI method is applied to obtain a deterministic result.
In conventional FWI, the gradient is computed using the adjoint method, and the loss function (\ref{l2 misfit}) is minimized via gradient descent.

\subsection{Synthetic data imaging with linear array transducer}

\begin{figure*}[ht]
	\centering
	\includegraphics[scale=0.4]{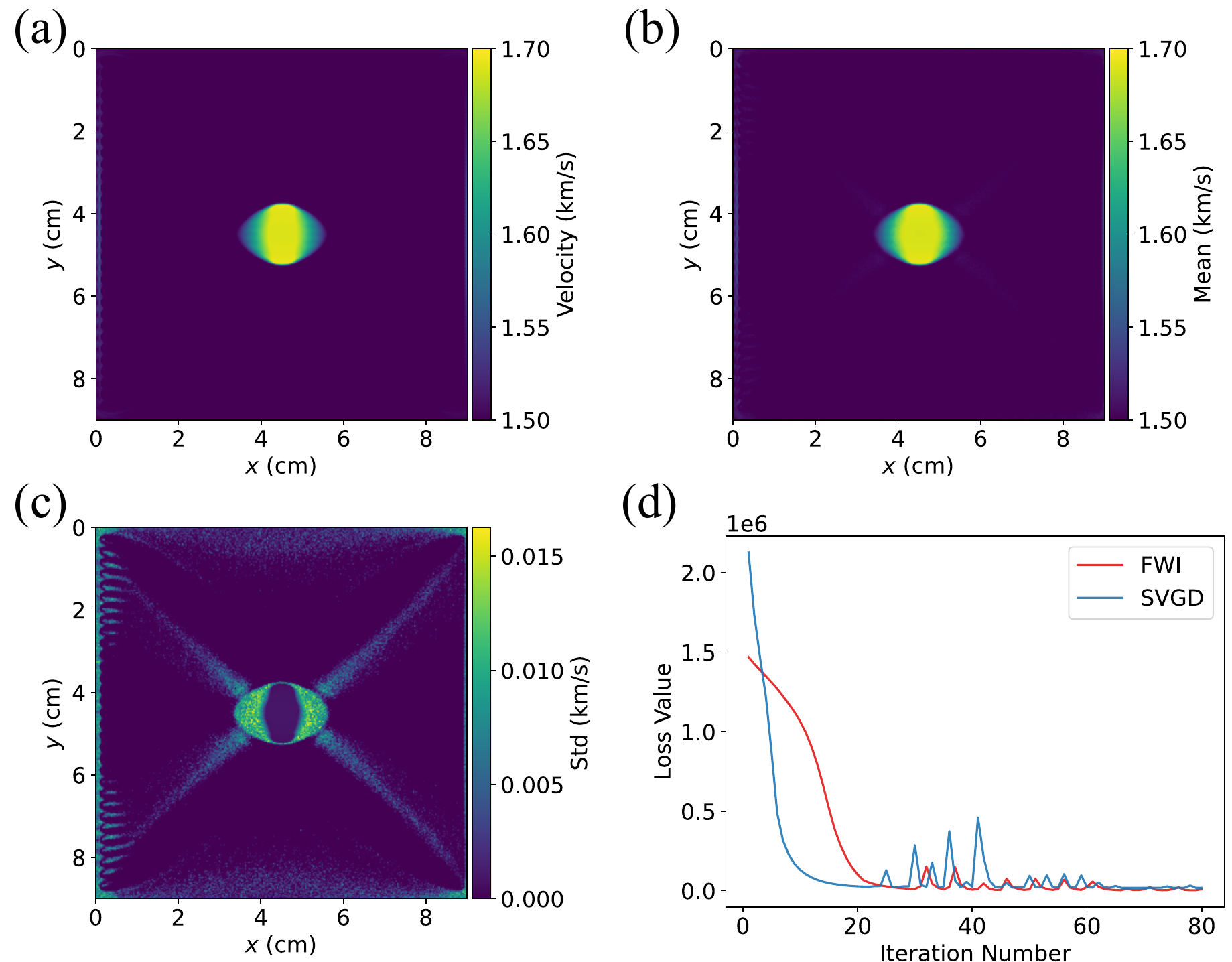}
	\caption{ Results of the inversion process for the linear array transducer simulation.
	(a) The SOS distribution obtained from the conventional FWI. 
	(b) The mean result of the SVGD particles.
	(c) The standard deviation of the SOS distribution from SVGD.
	(d) The loss values plotted against the iteration number. 
	}
	\label{fig:fwi2DlineSvgdRes}
\end{figure*}

First, the focus is on using synthetic data from a linear array transducer for inversion imaging. 
A two-dimensional grid model is employed, with a grid size of $ 301 \times 301$ and a spatial step size of $\Delta x=\Delta y= 0.3 $ mm, resulting in a total computational area of $ 9 \, \text{cm} \times 9 \, \text{cm} $. As shown in Figure \ref{fig:fwi2DlineSvgdInit}(a), the true model setup consists of a circular object located at the center of the computational area. The object is a homogeneous isotropic medium with a diameter of $ 1.5 $ cm and an SOS of $ 1.7 $ km/s. The background SOS is set at $ 1.5 $ km/s, resembling the speed of sound under water. 
This difference in SOS results in variations in the wave propagation, providing a basis for the inversion algorithm to distinguish the target object from the background medium.

The simulated experiment utilizes 31 transmitting sources and 101 receivers. In Figure \ref{fig:fwi2DlineSvgdInit}(a), the sources are evenly distributed at $x=0.15$ cm, and the receivers are uniformly placed at $x=8.9$ cm. 
The starting and ending positions of both sources and receivers are $0.3$ cm away from the upper and lower boundaries of the computational area, respectively. 
The source signal is a Ricker wavelet with a central frequency of 0.5 MHz and a duration of 0.12 ms. 
As illustrated in Figure \ref{fig:fwi2DlineSvgdInit}(b), the frequency spectrum of the Ricker wavelet is well-concentrated, delivering high energy within a narrow band.
The time step $\Delta t$ is determined based on the Courant-Friedrichs-Lewy (CFL) stability condition, $\Delta t \leq C \Delta x / c_{\text{max}}$, where $c_{\text{max}}$ is the maximum speed of sound in the model and $C$ is a dimensionless constant \cite{suo2023application}. With $C = 0.46$, the time step is set to $\Delta t = 8 \times 10^{-5}$ ms, resulting in 1501 temporal samples. To ensure numerical accuracy and suppress dispersion, the spatial discretization satisfies a minimum of 10 grid points per wavelength.

For the conventional FWI, the initial model sets the SOS within the grid to $1.5$ km/s, as shown in Figure \ref{fig:fwi2DlineSvgdInit}(c),
resulting in an initial squared slowness of $\bm{m} = 1/\bm{c}^2 \approx 0.444~\text{s}^2/\text{km}^2$. 
The true SOS ranges from $c_{\min} = 1.5$ km/s to $c_{\max} = 1.7$ km/s, the squared slowness is constrained within $m_{\min} = 1/c_{\max}^2$ and $m_{\max} = 1/c_{\min}^2$ during the inversion.
Values of $\bm{m}$ exceeding these bounds are truncated to  prevent numerical issues such as NaN (not a number) during computation. 
The inversion is performed using the steepest descent optimization method with a maximum of 80 iterations. 
The step size is empirically scheduled to decrease with iterations, initially set to $9 \times 10^{-3}$, and subsequently reduced by a decay factor of 0.57 every 20 iterations.
The stopping criterion is satisfied when either the relative change in the objective function between consecutive iterations falls below $10^{-5}$ or the iteration count reaches 80.

For the SVGD algorithm, 20 initial particles are generated by perturbing the initial model of the conventional FWI with Gaussian white noise (mean 0, standard deviation 0.01).
Figure \ref{fig:fwi2DlineSvgdInit}(d) displays one example of the initial particles. 
The initial particles are iteratively updated according to the SVGD update rule  Eq.(\ref{svgd update rule}). 
A truncated Gaussian prior Eq.(\ref{gaussian prior}) is adopted for the squared slowness, with $\mu_i=0.444~\text{s}^2/\text{km}^2$ and $\sigma_i=0.1 ~\text{s}^2/\text{km}^2$.  
The optimization procedure also employs the steepest descent method, using a maximum of 80 iterations. 
The step size is initialized at $9 \times 10^{-3}$ and decays by a factor of 0.57 every 20 iterations.
The inversion terminates when the relative change in the objective function between iterations drops below $10^{-5}$ or when the iteration limit is reached.

Figure~\ref{fig:fwi2DlineSvgdRes} presents the results of the inversion process, where the reconstructed SOS distribution is obtained via $\bm{c} = 1/\sqrt{\bm{m}}$. The result of the conventional FWI is shown in Figure~\ref{fig:fwi2DlineSvgdRes}(a).
The circular object in the true model is reconstructed as an ellipse in the inversion result. This transformation is primarily due to insufficient angular coverage. In the case of the linear array transducer, the sources and receivers do not provide adequate angular coverage, limiting the acquisition of wave propagation data from multiple angles. As a result, the inversion algorithm is unable to accurately reconstruct the true shape of the target object. A similar transformation is observed in Figure \ref{fig:fwi2DlineSvgdRes}(b), which shows the mean SOS distribution of the SVGD particles. 
Compared to the conventional FWI results, the mean SOS in SVGD particles is closer to the true model. 
The standard deviation, shown in Figure \ref{fig:fwi2DlineSvgdRes}(c), measures the uncertainty in the mean SOS distribution. 
The central region of Figure \ref{fig:fwi2DlineSvgdRes}(c) exhibits low uncertainty, indicating that the SVGD algorithm provides a more reliable SOS estimate in this area. 
Thus, the standard deviation serves to evaluate the accuracy and reliability of the mean result across different regions. 
Figure \ref{fig:fwi2DlineSvgdRes}(d) plots the loss values against the iteration number, showing the convergence of the inversion process. 

\subsection{Synthetic data imaging with ring array transducer}

\begin{figure*}[ht]
	\centering
	\includegraphics[scale=0.4]{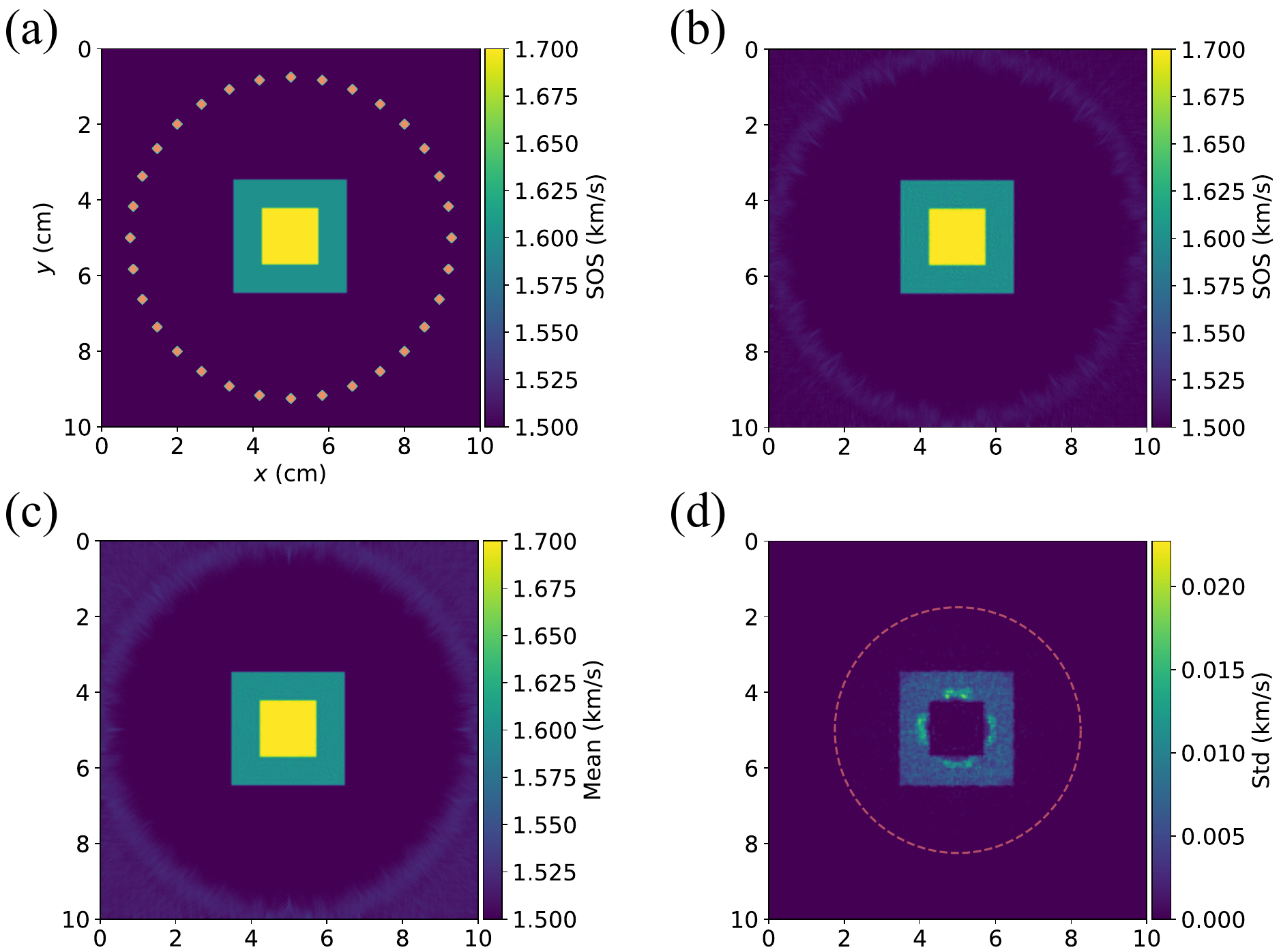}
	\caption{ Results of the inversion process for the ring array transducer simulation.
		(a) The true model setup. The orange dots and green diamonds indicate the locations of the sources and receivers, respectively.
		(b) The SOS distribution obtained from the conventional FWI. 
		(c) The mean SOS estimated by SVGD using 20 particles.
		(d) The standard deviation of the SOS distribution from SVGD.
		The pink dashed circle (radius = 3.27 cm) indicates the ROI used to evaluate the effectiveness of the inversion algorithms.
		}
	\label{fig:fwi2DringSvgd}
\end{figure*}

To mitigate the limitations of insufficient angular coverage inherent in linear array transducer configurations, the experiment adopts a ring array layout commonly employed in medical ultrasound imaging. 
A two-dimensional grid model comprising $335 \times 335$ points with a spatial resolution of $\Delta x = \Delta y = 0.3$ mm defines a computational domain measuring $10.02 \, \text{cm} \times 10.02 \, \text{cm}$. 
As depicted in Figure~\ref{fig:fwi2DringSvgd}(a), the true model includes a central rectangular object with dimensions of 3 cm × 3 cm. The innermost core, measuring 1.5 cm × 1.5 cm, has a SOS of 1.7 km/s, surrounded by a region with an SOS of 1.6 km/s, embedded in a background with an SOS of 1.5 km/s. 
A total of 32 sources and 32 receivers are evenly distributed along a circular ring with a diameter of 8.52 cm, centered within the grid. The receiver layout mirrors that of the sources.  The excitation signal is a Ricker wavelet with a central frequency of 0.5 MHz and a duration of 0.12 ms. To ensure numerical stability, the time step $\Delta t$ adheres to the CFL condition with $C = 0.3$, resulting in $\Delta t = 8 \times 10^{-5}$ ms and a total of 1601 temporal samples. The spatial discretization satisfies the requirement of at least 10 grid points per wavelength to minimize numerical dispersion.

For the conventional FWI, the initial model sets the SOS uniformly to $1.5$ km/s (the approximate SOS of pure water), resulting in an initial squared slowness of $\bm{m} = 1/\bm{c}^2 \approx 0.444~\text{s}^2/\text{km}^2$. The true SOS varies within the range $c_{\min} = 1.5$ km/s to $c_{\max} = 1.7$ km/s, and the squared slowness is accordingly constrained within $m_{\min} = 1/c_{\max}^2$ and $m_{\max} = 1/c_{\min}^2$ during the inversion. Any values of $\bm{m}$ that fall outside this range are truncated to ensure physical plausibility and to prevent numerical issues such as NaN during computation. The inversion is performed using the quasi-Newton optimization method L-BFGS-B, implemented via the scipy.optimize.minimize function in Python, with a maximum of 50 iterations. The algorithm employs an internal line search based on strong Wolfe conditions, with the step length adaptively chosen. The stopping criterion is met when the relative change in the objective function falls below $10^{-5}$ or when the number of iterations reaches the specified maximum.

For the SVGD algorithm, 20 initial particles are generated by adding Gaussian white noise (mean 0, standard deviation 0.01) to the initial FWI model. The particles are iteratively updated according to Eq. (\ref{svgd update rule}). 
The prior distribution is modeled as a truncated Gaussian distribution Eq.(\ref{gaussian prior}), with $\mu_i = 0.444~\text{s}^2/\text{km}^2$ and $\sigma_i = 0.1~\text{s}^2/\text{km}^2$. 
The optimization is carried out using the quasi-Newton method L-BFGS-B, implemented via the scipy.optimize.minimize function, with a maximum of 50 iterations. The algorithm incorporates an internal line search, and the step size is adaptively selected. The inversion terminates when the relative change in the objective function falls below $10^{-5}$ or when the iteration count reaches the maximum limit.

\begin{figure*}[ht]
	\centering
	\includegraphics[width=\linewidth]{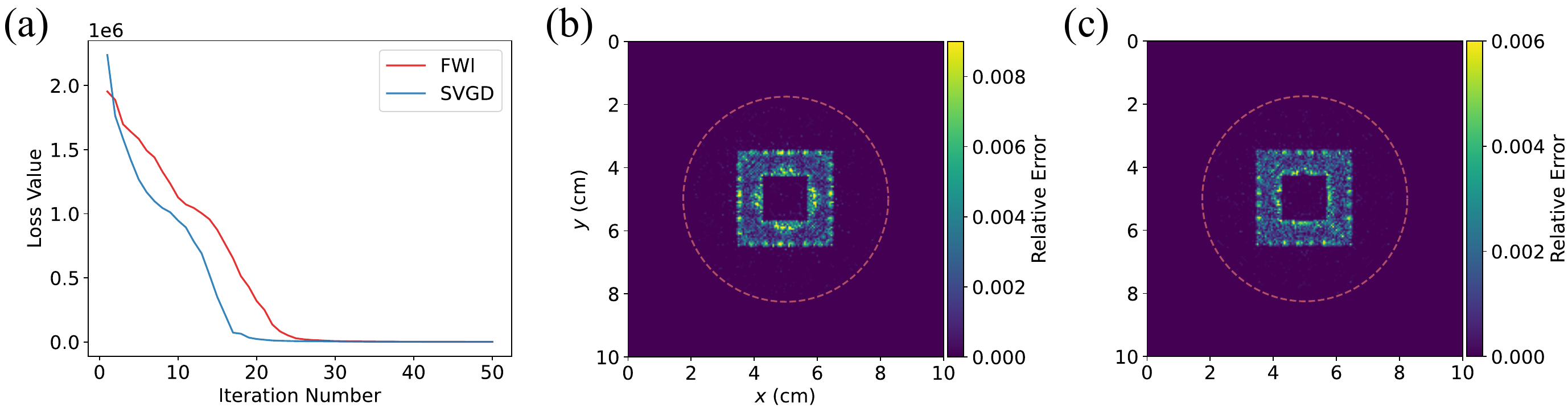}
	\caption{Loss and relative error analysis for the ring array transducer simulation. 
		(a) The loss values plotted against the iteration number. 
		(b) and (c) show the relative error for FWI and SVGD, respectively.
		The pink dashed circle (radius = 3.27 cm) indicates the ROI.
	}
	\label{fig:fwi2DringLossErr}
\end{figure*}

The reconstructed SOS is obtained by $\bm{c} = 1/\sqrt{\bm{m}}$. For the SVGD method, $n$ particles ($n=20$) produce $n$ distinct reconstructions $\bm{c}_i,~i=1,2,\cdots,n$, from which the mean and standard deviation of the SOS distribution are computed. 
Figure \ref{fig:fwi2DringSvgd}(b)-(c) show the SOS distribution obtained from the conventional FWI and SVGD, respectively. 
Both methods accurately reconstruct the true model, capturing the internal structure and contours of the object.
Figure \ref{fig:fwi2DringSvgd}(d) measures the imaging uncertainty of the SVGD mean SOS. High uncertainty appears mainly in the green region of Figure \ref{fig:fwi2DringSvgd}(c). This area shows subtle SOS variations around 1.6 km/s, which also appear in the conventional FWI result Figure \ref{fig:fwi2DringSvgd}(b). 

Figure \ref{fig:fwi2DringLossErr}(a) shows the loss values of FWI and SVGD as a function of iteration number, with SVGD converging more quickly.
To quantify reconstruction accuracy, the relative error between the predicted model $\hat{\bm{m}}$ and the true model $\bm{m}$, defined as $(\hat{\bm{m}} - \bm{m}) / \bm{m}$, is computed.
Figures \ref{fig:fwi2DringLossErr}(b) and (c) display the spatial distribution of the relative error for FWI and SVGD, respectively. Within the ROI (pink dashed circle), SVGD achieves a lower maximum relative error of 0.84\% and a mean error of 0.035\%, compared to 1.563\% and 0.064\% for FWI.
The results suggest that the SVGD algorithm performs well in inversion imaging, and provides an additional standard deviation map to evaluate the voxel-wise uncertainty of the SOS estimate.

\begin{figure*}[ht]
	\centering
	\includegraphics[width=\linewidth]{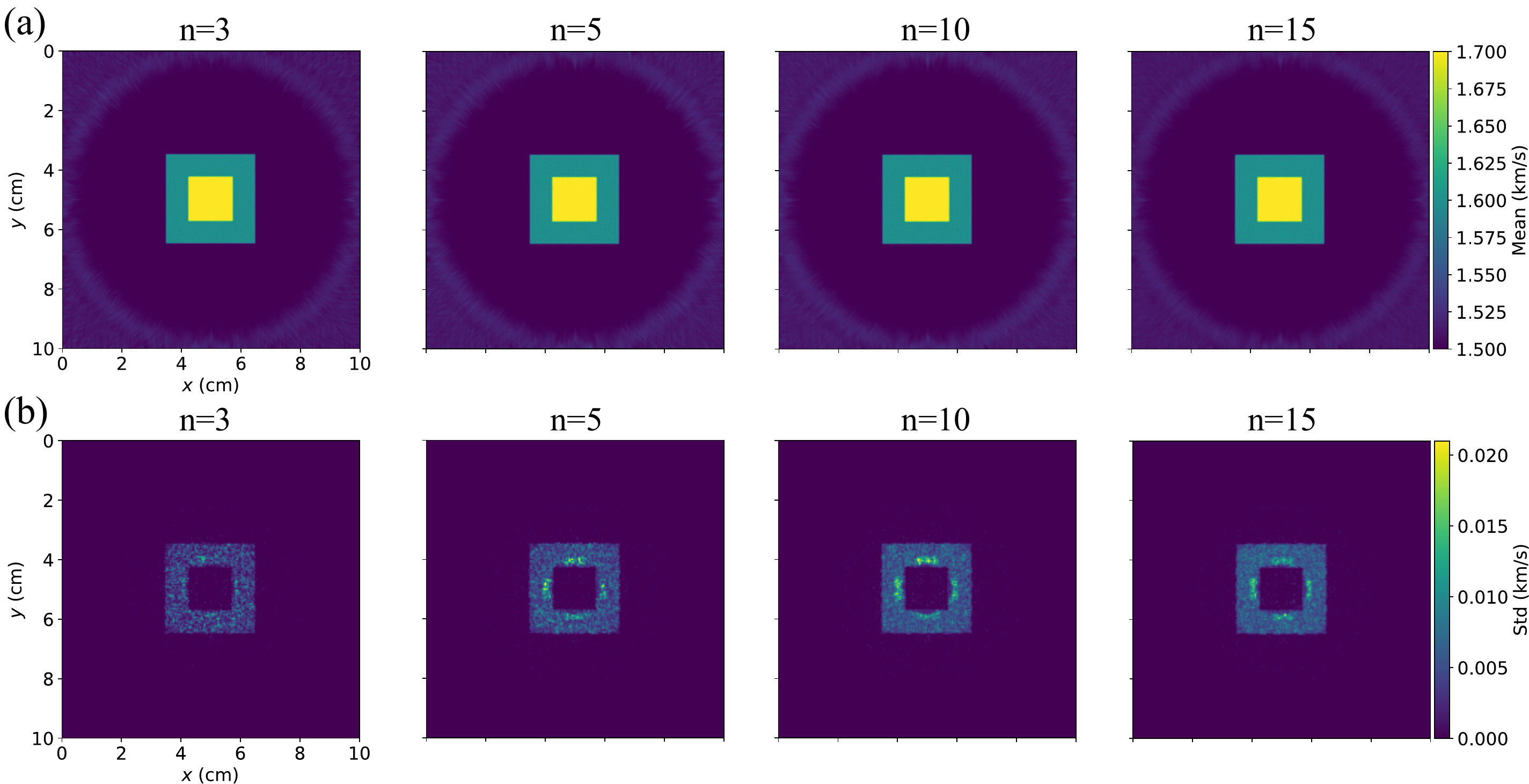}
	\caption{ Sensitivity of the SVGD algorithm to the number of particles.
		(a) The mean SOS for different particle counts ($n = 3, 5, 10, 15$, from left to right).
		(b) The standard deviation of the SOS.
	}
	\label{fig:fwi2DringSvgdNum_1}
\end{figure*}

In SVGD inversion, one important hyperparameter is the number of particles, denoted as $n$. Figure \ref{fig:fwi2DringSvgdNum_1} (a) and (b) show the mean and the standard deviation of SOS, respectively, for different particle counts: $n = 3, 5, 10, 15$ (from left to right). It can be observed that even with a small number of particles, the mean SOS inversion using the SVGD algorithm still produces satisfactory results. As the number of particles increases, the standard deviation distribution becomes more consistent, with the boundaries between different sound speeds becoming clearer. Additionally, for higher particle counts, the standard deviation at the low-to-high speed boundaries within the tissue becomes larger. This could potentially reflect the reflection or scattering of sound waves at these boundaries, highlighting the improved precision of the inversion with more particles.

\begin{figure*}[ht]
	\centering
	\includegraphics[width=\linewidth]{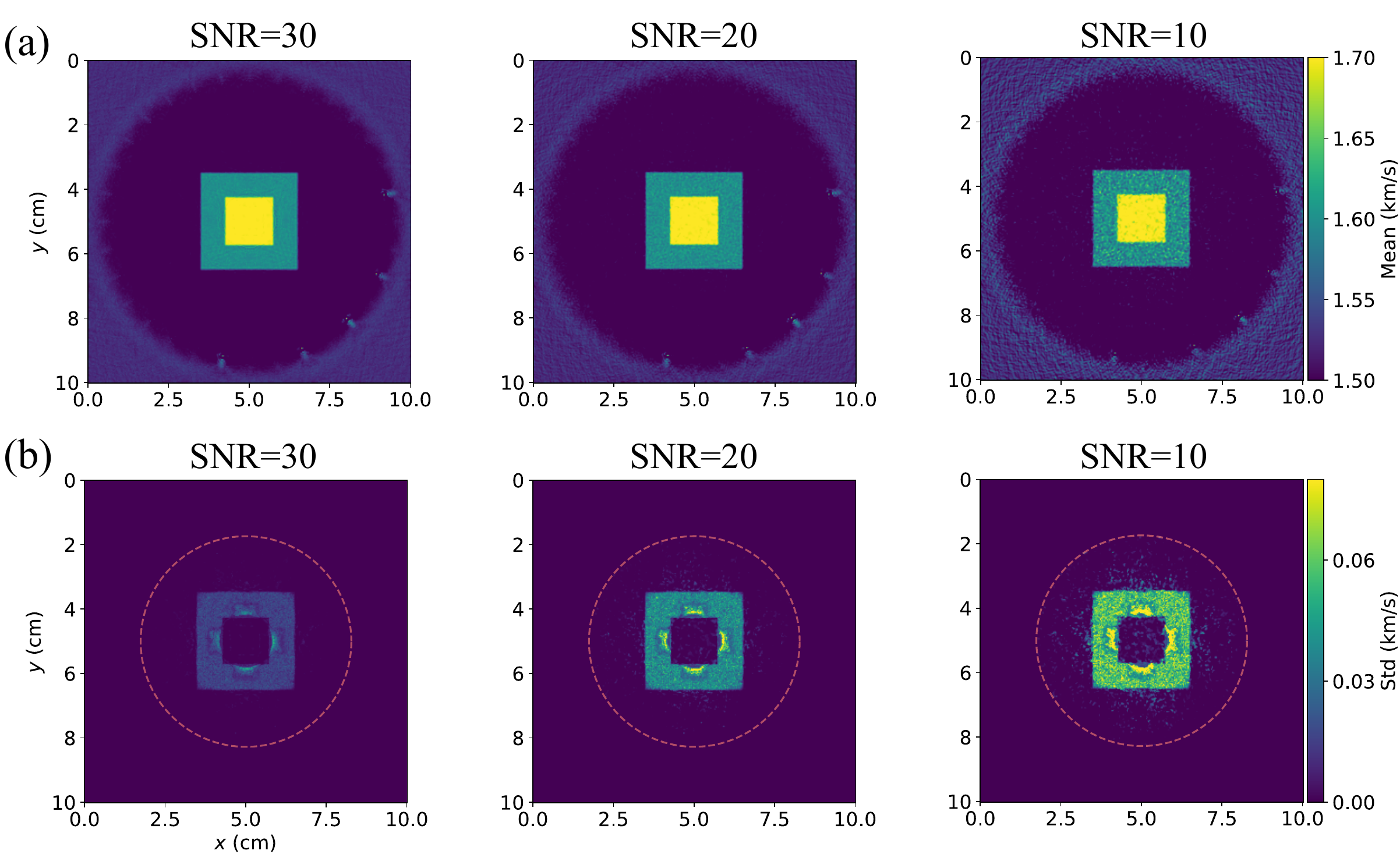}
	\caption{ Uncertainty quantification of SVGD-based inversion under varying noise levels. 
		(a) Mean SOS.
		(b) Standard deviation of the SOS. 
		The inversion employs 20 SVGD particles. 
		The left, middle, and right panels correspond to SNRs of 30~dB, 20~dB, and 10~dB, respectively.
		The pink dashed circle (radius = 3.27 cm) indicates the ROI.
	}
	\label{fig:fwi2dNoiseSVGDvp}
\end{figure*}

To further assess the uncertainty quantification capability of the SVGD-based inversion method, we investigate the variation of posterior uncertainty under different levels of measurement noise.
Gaussian noise with zero mean is added to the synthetic observed data, with signal-to-noise ratios (SNRs) of 30 dB, 20 dB, and 10 dB representing low, moderate, and high noise levels, respectively.
For each noise level, the inversion is performed using 20 SVGD particles, initialized by adding Gaussian white noise (mean 0, standard deviation 0.01) to the initial pure  water model.
The prior distribution is modeled as a truncated Gaussian distribution, with $\mu_i = 0.444~\text{s}^2/\text{km}^2$ and $\sigma_i = 0.1~\text{s}^2/\text{km}^2$.
The uncertainty is evaluated by computing the pointwise standard deviation across the particles. 
As shown in Figure \ref{fig:fwi2dNoiseSVGDvp}(a), the mean of SOS remains relatively accurate under low and moderate noise, while performance deteriorates under high noise.
The corresponding standard deviation maps in Figure \ref{fig:fwi2dNoiseSVGDvp}(b) indicate increased uncertainty with decreasing SNR.
The results demonstrate that SVGD captures the relationship between measurement noise and uncertainty in a consistent manner.

%Specifically, within the ROI (pink dashed circle), the maximum standard deviations are 0.06, 0.14, and 0.24, and the average standard deviations are 0.003, 0.008, and 0.013 for 30 dB, 20 dB, and 10 dB, respectively. 

\subsection{Breast tissue imaging using SVGD and FWI algorithms}

\begin{figure*}[ht]
	\centering
	\includegraphics[scale=0.4]{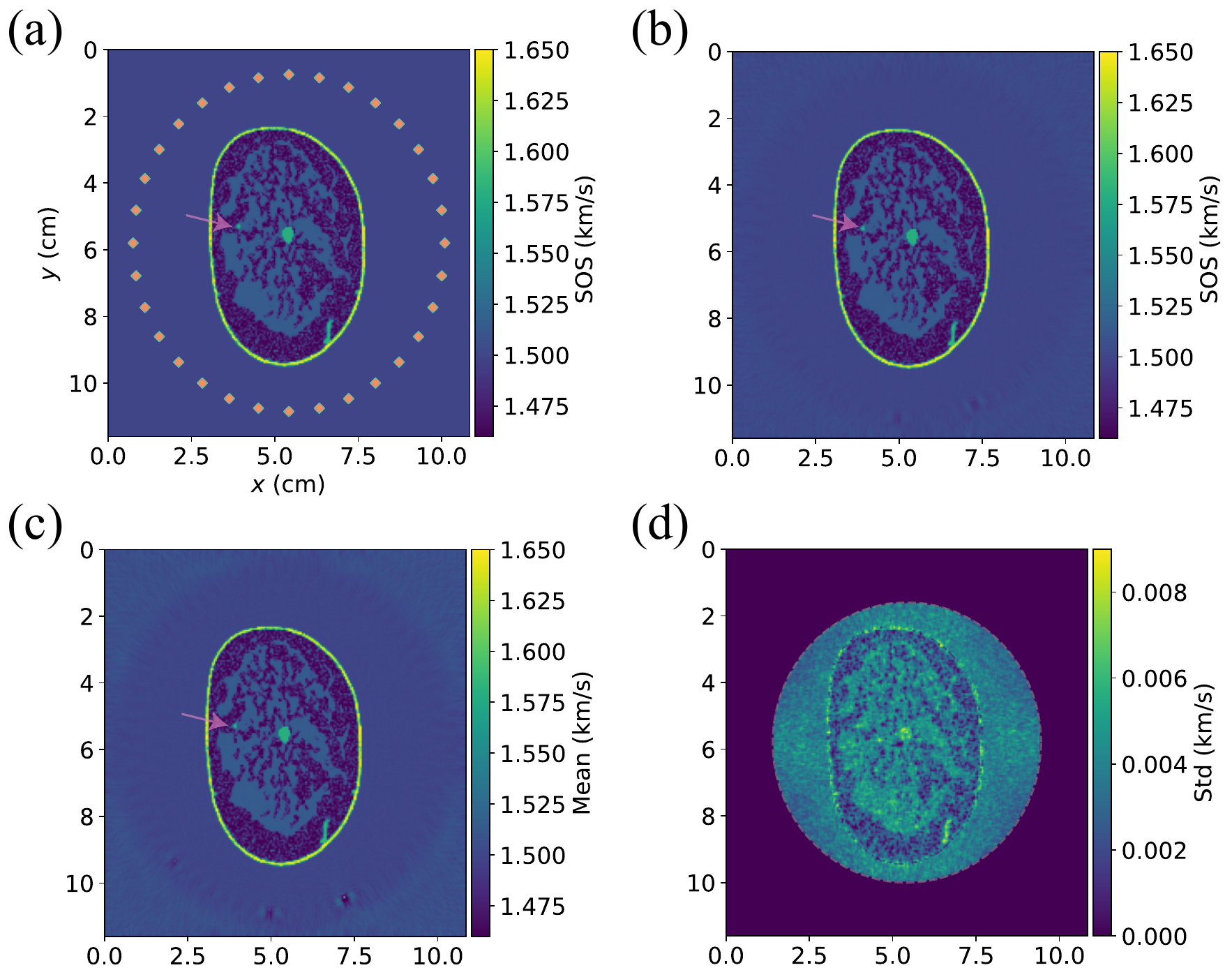}
	\caption{Inversion results for the breast tissue model with a tumor.
	(a) The true SOS distribution in breast tissue, incorporating acoustic tissue properties and a tumor. 
	The orange dots and green diamonds indicate the locations of the sources and receivers, respectively.
	(b) The SOS distribution obtained from the conventional FWI. 
	(c) The mean result of the SVGD algorithm using 20 particles.
	The magenta arrows in (a)-(c) highlight the location of the tumor.
	(d) The standard deviation of the SOS distribution from the SVGD algorithm.
	}
	\label{fig:fwi2DbreastSvgd}
\end{figure*}

In the previous paragraphs, the effectiveness of the SVGD algorithm is validated using both linear and ring array transducers. 
To further assess the performance under complex anatomical structures, we apply SVGD to a simulated breast tissue model derived from an open database.
The model incorporates realistic acoustic tissue properties and includes an embedded tumor, as described in \cite{lou2017generation, cueto2022stride}.
Figure \ref{fig:fwi2DbreastSvgd}(a) shows the true SOS model. 
The imaged object represents the SOS distribution of breast tissue, with SOS values ranging from 1.46 km/s to 1.65 km/s. 
The computational domain is discretized into $356 \times 385$ grid points with a uniform spatial spacing of 0.3 mm. A total of 64 sources and 64 receivers are uniformly arranged in an elliptical pattern centered within the domain. The transmitted waveform is a Ricker wavelet with a central frequency of 0.5 MHz and a duration of 0.09 ms. The time step $\Delta t$ is determined based on the CFL condition, using the stability constant $C = 0.3$, resulting in $\Delta t = 5 \times 10^{-5}$ ms and a total of 1801 temporal samples. The spatial discretization corresponds to approximately 10 grid points per wavelength.

\begin{figure*}[ht]
	\centering
	\includegraphics[width=\linewidth]{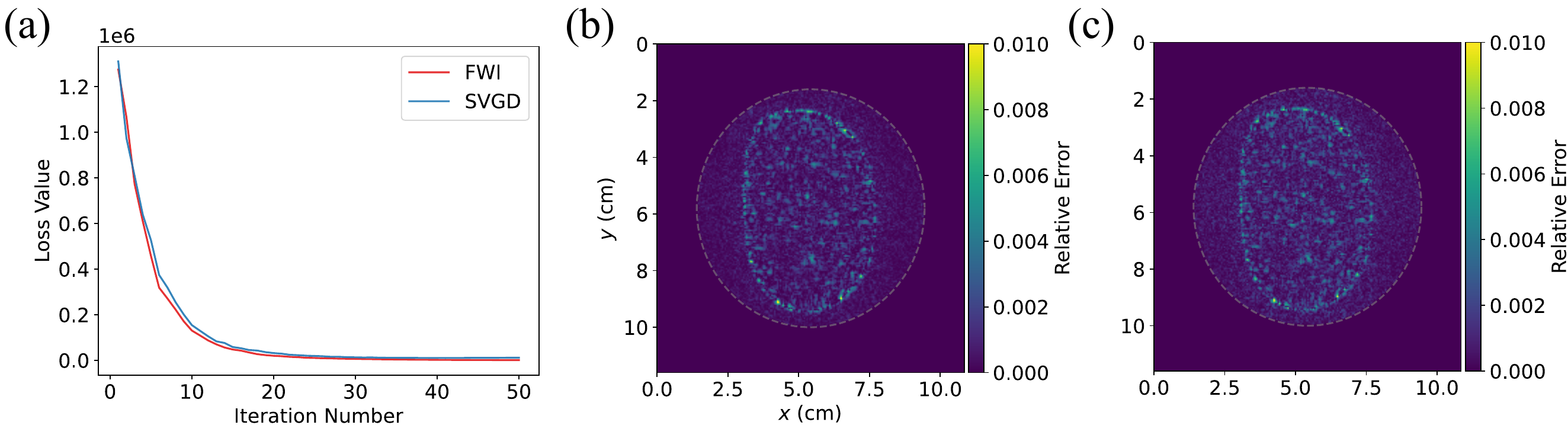}
	\caption{Loss and relative error analysis for breast tissue inversion imaging. 
	(a) The loss values plotted against the iteration number. 
	(b) and (c) show the relative error between the predicted and true SOS for FWI and SVGD, respectively. The gray dashed ellipse highlights the region to assess the effectiveness of the inversion algorithms.
	}
	\label{fig:fwi2DbreastLossErr}
\end{figure*}

In the inversion process, the initial model for conventional FWI is set to a constant SOS of $1.5$ km/s, resulting in an initial squared slowness of $\bm{m} = 1/\bm{c}^2 \approx 0.444~\text{s}^2/\text{km}^2$. The true SOS varies within the range $c_{\min} = 1.46$ km/s to $c_{\max} = 1.65$ km/s, and the squared slowness is constrained within $m_{\min} = 1/c_{\max}^2$ and $m_{\max} = 1/c_{\min}^2$ during the inversion. The optimization is performed using the quasi-Newton method L-BFGS-B, implemented via the scipy.optimize.minimize function in Python, with a maximum of 50 iterations. The algorithm utilizes an internal line search procedure in which the step length is adaptively determined at each iteration. The stopping criterion is satisfied when the relative change in the objective function falls below $10^{-5}$ or when the maximum number of iterations is reached.

The SVGD initialization involves generating 20 particles by injecting Gaussian noise (mean 0, standard deviation 0.01) into the initial FWI model.
The prior distribution follows the truncated Gaussian distribution (\ref{gaussian prior}) with $\mu_i = 0.444~\text{s}^2/\text{km}^2$ and $\sigma_i = 0.1~\text{s}^2/\text{km}^2$. 
The optimization is performed using the quasi-Newton method L-BFGS-B, with a maximum of 50 iterations. 
The algorithm employs an internal line search procedure, and the step length is adaptively determined. 
The stopping criterion is met when the objective change is below $10^{-5}$ or the iteration limit is reached.

%The inversion process uses a Ricker wavelet with a fixed central frequency of 0.5 MHz.
Figures \ref{fig:fwi2DbreastSvgd}(b) and (c) show the SOS distributions obtained by FWI and SVGD, respectively. Both methods successfully reconstruct the breast tissue information, identifying structures such as the skin, fat, stroma, and tumor. Figure \ref{fig:fwi2DbreastSvgd}(d) measures the imaging uncertainty of the SVGD mean SOS. The uncertainty varies across different tissues in the breast, with most values smaller than 0.01 and a mean uncertainty of approximately 0.003. The low uncertainty suggests that the SVGD mean SOS provides a reliable estimate of the breast tissue properties.

Figure \ref{fig:fwi2DbreastLossErr}(a) shows the loss values plotted against the iteration number for both the conventional FWI and SVGD algorithms.
The FWI algorithm reaches a lower loss value after 50 inversion steps, which is due to the different loss functions used by FWI and SVGD. FWI minimizes the $l_2$ data misfit, while SVGD minimizes the KL divergence between the unnormalized posterior and the inferred distribution. 
Figures \ref{fig:fwi2DbreastLossErr}(b) and (c) display the relative error between the predicted and true SOS for FWI and SVGD, respectively. Within the gray dashed ellipse, the maximum relative error for FWI is $1.25\%$, with a mean relative error of $0.08\%$. For SVGD, the maximum relative error is $1.10\%$, and the mean relative error is $0.09\%$. 
Both FWI and SVGD algorithms provide accurate and reliable estimates of the breast tissue SOS distribution.

\begin{figure*}[ht]
	\centering
	\includegraphics[width=\linewidth]{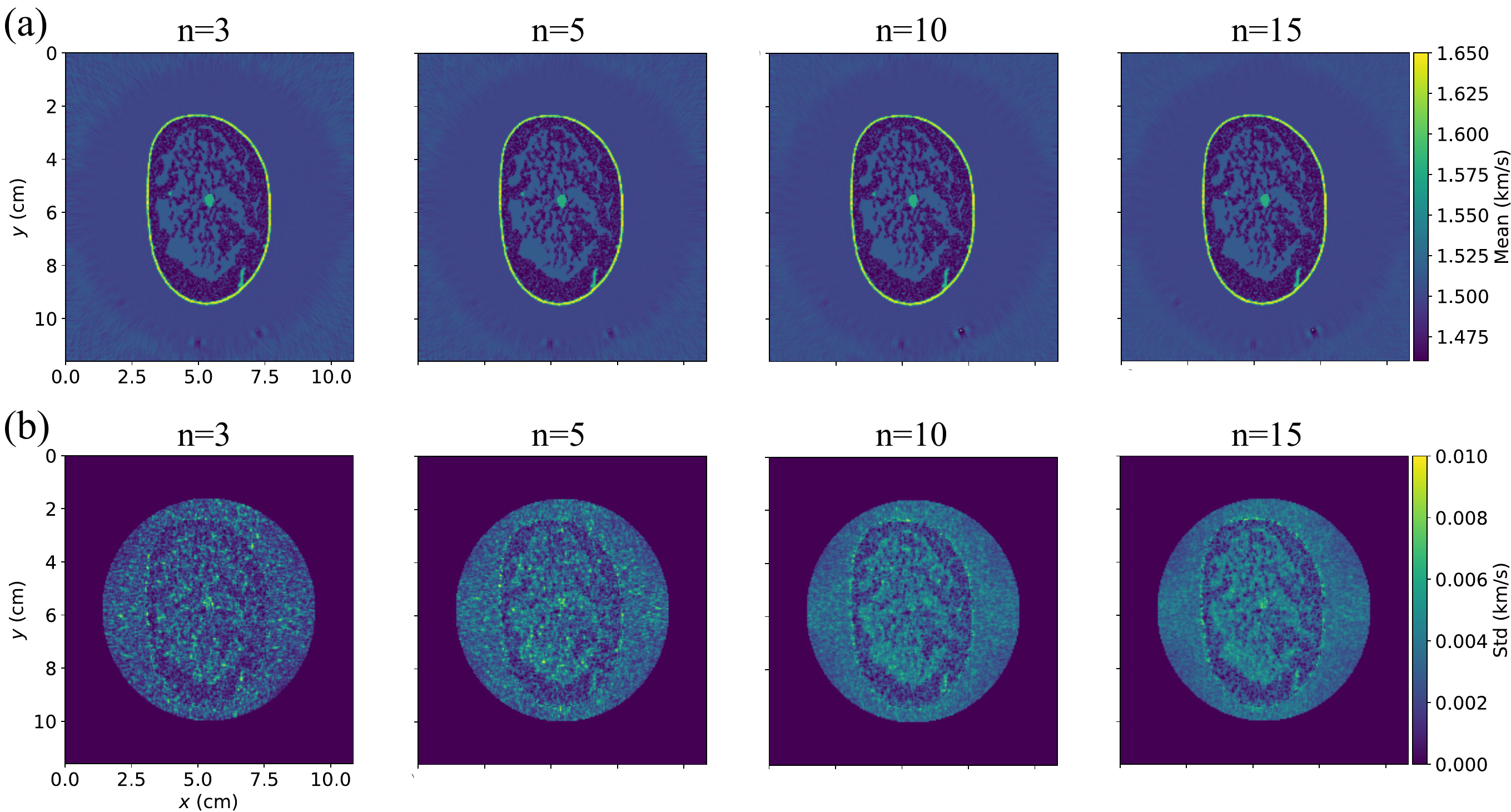}
	\caption{ Sensitivity of the SVGD algorithm to particle count in breast tissue data.
		(a) Mean SOS for $n = 3, 5, 10, 15$ (from left to right).
		(b) Standard deviation of the SOS distribution for each particle count.
	}
	\label{fig:fwi2DbreastSvgdNum_1}
\end{figure*}

Figure \ref{fig:fwi2DbreastSvgdNum_1} demonstrates the sensitivity of the SVGD algorithm to the number of particles used in the inversion process. Panels (a) and (b) show the mean and standard deviation of the SOS, respectively, for particle counts of $n = 3, 5, 10, 15$ (from left to right). Even with a small number of particles, the SVGD algorithm produces reliable mean SOS results. As the number of particles increases, the standard deviation distribution becomes more refined, with clearer boundaries between different tissue types. 
The results suggest the advantages of the SVGD algorithm, demonstrating that the method is scalable by adjusting the particle count. Even with fewer particles, the algorithm still achieves satisfactory inversion results, highlighting its efficiency and flexibility in handling different particle sizes.

The results in Figures \ref{fig:fwi2DbreastSvgd}-\ref{fig:fwi2DbreastLossErr} validate the effectiveness of using the mean value from SVGD for reconstructing breast tissue. 
To further assess the quality of the obtained standard deviation, 
we conduct a comparative analysis of SVGD against two Bayesian inference methods,
namely SVI \cite{bates2022probabilistic} and Metropolis-Hastings MCMC \cite{xie2024stochastic,izzatullah2021bayesian}.
This comparison enables a more rigorous evaluation of SVGD's capability for uncertainty quantification.

\subsection{Uncertainty Quantification via SVI-Based FWI and Metropolis-Hastings MCMC}
\begin{figure*}[ht]
	\centering
	\includegraphics[scale=0.4]{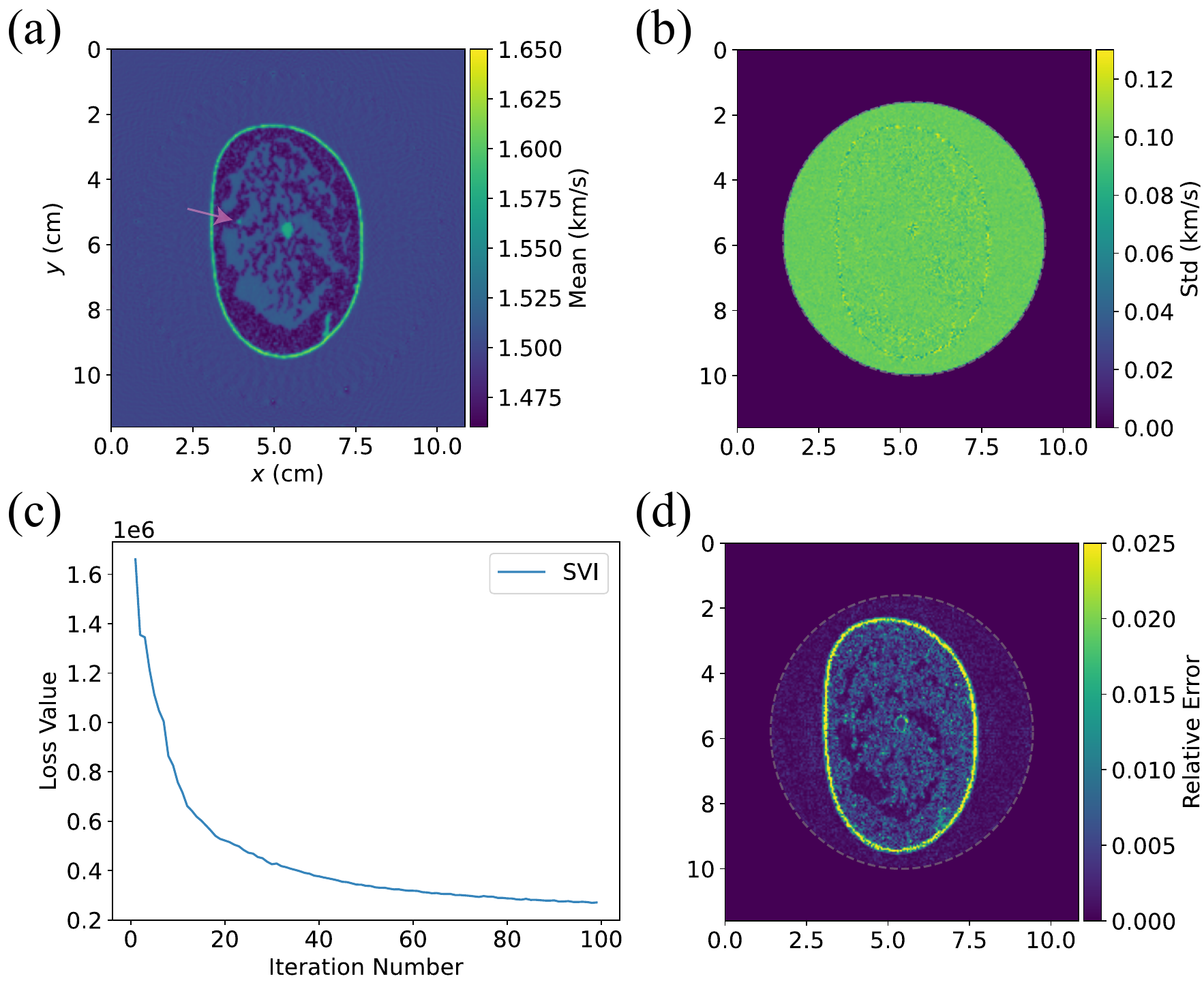}
	\caption{SVI results for breast tissue.
		(a) The mean SOS from the SVI algorithm, with the magenta arrow highlighting the location of the tumor.
		(b) The standard deviation of the SOS distribution.
		(c) The loss values plotted against the iteration number.
		(d) The relative error between the predicted and true SOS. The gray dashed ellipse highlights the region to assess the effectiveness of the inversion algorithms.
	}
	\label{fig:fwi2DbreastSviRes}
\end{figure*}

The SVI-based FWI is a probabilistic approach that leverages stochastic variational inference to approximate the posterior distribution of SOS. 
SVI assumes a mean-field Gaussian prior that is spatially uncorrelated, with an initial mean $ \tilde{\bm \mu} $ and standard deviation $ \tilde{\bm \sigma} $. To enhance the accuracy and stability of gradient estimation, the method draws $ n $ samples $ \bm{c}_i^t $ at each iteration $ t $ from the current Gaussian prior $ \mathcal{N}(\tilde{\bm \mu}^t, \tilde{\bm \sigma}^t) $. The mean and standard deviation of the prior are updated by~\cite{bates2022probabilistic}
\begin{equation} \label{svi update rule}
	\tilde{\bm \mu}^{t+1} = \tilde{\bm \mu}^t + \frac{1}{n} \sum_{i=1}^{n} \mathbf{\Delta}_i^t, ~~
	\tilde{\bm \sigma}^{t+1} = \tilde{\bm \sigma}^t + \frac{1}{n}\sum_{i=1}^{n} \bm{\eta}_i^t \cdot \mathbf{\Delta}_i^t,
\end{equation}
where $ \bm{\eta}_i^t \sim \mathcal{N}(0, \mathbf{I}) $ introduces stochastic perturbations. The term $ \mathbf{\Delta}_i^t $ denotes the adjoint-state gradient of the L2-norm data misfit with respect to sample $ \bm{c}_i^t $. Upon convergence or after a predefined number of iterations, the mean value is used to reconstruct the tissue properties, and the standard deviation quantifies voxel-wise uncertainty.

For the SVI-based inversion, the initial prior distribution is modeled as a truncated Gaussian with mean $\tilde\mu_i^0 = 1.5~\text{km}/\text{s}$ and standard deviation $\tilde\sigma_i^0 = 0.1~\text{km}/\text{s}$. The prior distribution is iteratively refined using the SVI update rule Eq.~(\ref{svi update rule}). At each iteration, 20 samples $\bm c_i^t$ are generated from the current prior distribution. 
The optimization follows a steepest descent scheme with a maximum of 100 iterations. The step size is initially set to $5 \times 10^{-3}$ and is reduced by a factor of 0.8 every 20 iterations. The inversion terminates when either the relative change in the objective function falls below $10^{-5}$ or the iteration limit is reached.

Figure \ref{fig:fwi2DbreastSviRes}(a) presents the mean SOS obtained from the SVI method, where both the breast tissue and the tumor are well reconstructed. 
However, the standard deviation shown in Figure \ref{fig:fwi2DbreastSviRes}(b) fails to effectively quantify the uncertainty in the imaging results.
In comparison to the uncertainty illustrated in Figure \ref{fig:fwi2DbreastSvgd}(d), only the skin layer is clearly depicted in Figure \ref{fig:fwi2DbreastSviRes}(b). 
The maximum standard deviation within the ROI is 0.14, with an average of 0.1, which is larger than that obtained from SVGD. 
From the variations in loss values over iterations shown in Figure \ref{fig:fwi2DbreastSviRes}(c), it is evident that the SVI method achieves the converged SOS results. Figure \ref{fig:fwi2DbreastSviRes}(d) shows the relative error between the predicted and true SOS for SVI. 
Inside the gray dashed circle, the maximum relative error is $5.054\%$, and the mean relative error is $0.396\%$.
The relative error of SVI is larger than that of conventional FWI and SVGD methods.

The standard deviation of the posterior distribution reflects the intrinsic uncertainty of inverse problems. It primarily depends on the choice of the prior distribution and the level of observational noise. 
Therefore, different algorithms should not produce significantly different estimates of posterior uncertainty. To provide a reliable reference for uncertainty quantification, we employ the Metropolis-Hastings MCMC (MH-MCMC) method to sample from the posterior distribution of the model parameters.

The MH-MCMC algorithm constructs a Markov chain whose stationary distribution converges to the target posterior distribution after sufficient iterations. 
The algorithm starts by drawing a candidate sample $\bm{m}'$ from a proposal distribution $q(\bm{m}' \mid \bm{m}^t)$, which denotes the probability of $\bm{m}'$ given the current state $\bm{m}^t$.
The candidate is accepted with probability $\alpha$, defined as
\begin{equation} \label{alaph define}
	\alpha = \min\left\{1, \frac{p(\bm{m}' \mid \bm{d}^{\text{obs}})\, q(\bm{m}^{t} \mid \bm{m}')}
	{p(\bm{m}^{t} \mid \bm{d}^{\text{obs}})\, q(\bm{m}' \mid \bm{m}^{t})} \right\},
\end{equation}
where $p(\bm{m}^t \mid \bm{d}^{\text{obs}})$ denotes the posterior probability, and  $q(\bm{m}^{t} \mid \bm{m}')$ is the reverse proposal distribution. 
If accepted, the Markov chain moves to the candidate state, i.e., $\bm{m}^{t+1} = \bm{m}'$. 
Otherwise, it remains at the current state, $\bm{m}^{t+1} = \bm{m}^t$. This process is repeated until a specified number of iterations is reached or convergence is achieved.

To enhance sampling efficiency and acceptance rate, the Metropolis-adjusted Langevin Algorithm (MALA) is incorporated into the MH-MCMC framework.
MALA leverages gradient-informed proposals to improve sampling directionality and mitigate the inefficiency of random-walk behavior in traditional MH-MCMC algorithms. 
The proposal distribution for generating candidate samples $\bm{m}'$ is \cite{xie2024stochastic,izzatullah2021bayesian}
\begin{equation} \label{theta' update}
	\bm{m}' = \bm{m}^t + \frac{\epsilon_t^2}{2} \nabla \log p(\bm{m}^t \mid \bm{d}^{\text{obs}})  + \epsilon_t \bm{\eta}_t,
\end{equation}
where $\bm{\eta}_t \sim \mathcal{N}(0, \mathbf{I})$ is a $d$-dimensional standard Gaussian random variable.
Each update in Eq.~(\ref{theta' update}) combines a gradient-driven drift term, which pulls particles toward high-probability regions of the posterior, with a stochastic diffusion term that preserves exploration and ensures ergodicity. 
The step size $\epsilon_t$ determines how much influence the gradient term and the noise term have on the candidate sample.
This combination enables more efficient exploration of the posterior distribution in high-dimensional inverse problems.

In practical implementations of the MH-MCMC method, an initial set of samples, referred to as the burn-in period, is typically discarded to allow the Markov chain to converge to its stationary distribution.
To accelerate convergence and ensure high-quality initialization in posterior sampling of breast tissue data, a two-stage strategy is employed.
During the burn-in stage, 20 chains are initialized from a normal distribution with a mean of $1/1.5^2$ (assuming the SOS in pure water is 1.5 km/s) and a standard deviation of 0.01.
To emphasize the accurate estimation of posterior uncertainty, the prior distribution $p(\bm{m})$ is also the truncated Gaussian, centered at the reciprocal of the squared true SOS, with a standard deviation of 0.1. 
The MALA proposal Eq.~(\ref{theta' update}) is used to update each of the 20 chains.
The initial step size is set to $\epsilon_t = 1 \times 10^{-2}$, with a gradual reduction in step size over 100 iterations, and all samples from this stage are discarded to eliminate initialization bias \cite{xie2024stochastic}. 
In the subsequent MH-MCMC sampling stage, 100 iterations are performed using a smaller step size of $\epsilon_t = 2 \times 10^{-4}$.
At each iteration, a candidate sample is generated and evaluated using the Metropolis-Hastings acceptance probability $\alpha$.
If the candidate is accepted, it replaces the current sample; otherwise, the chain remains at the current state.

\begin{figure*}[ht]
	\centering
	\includegraphics[scale=0.4]{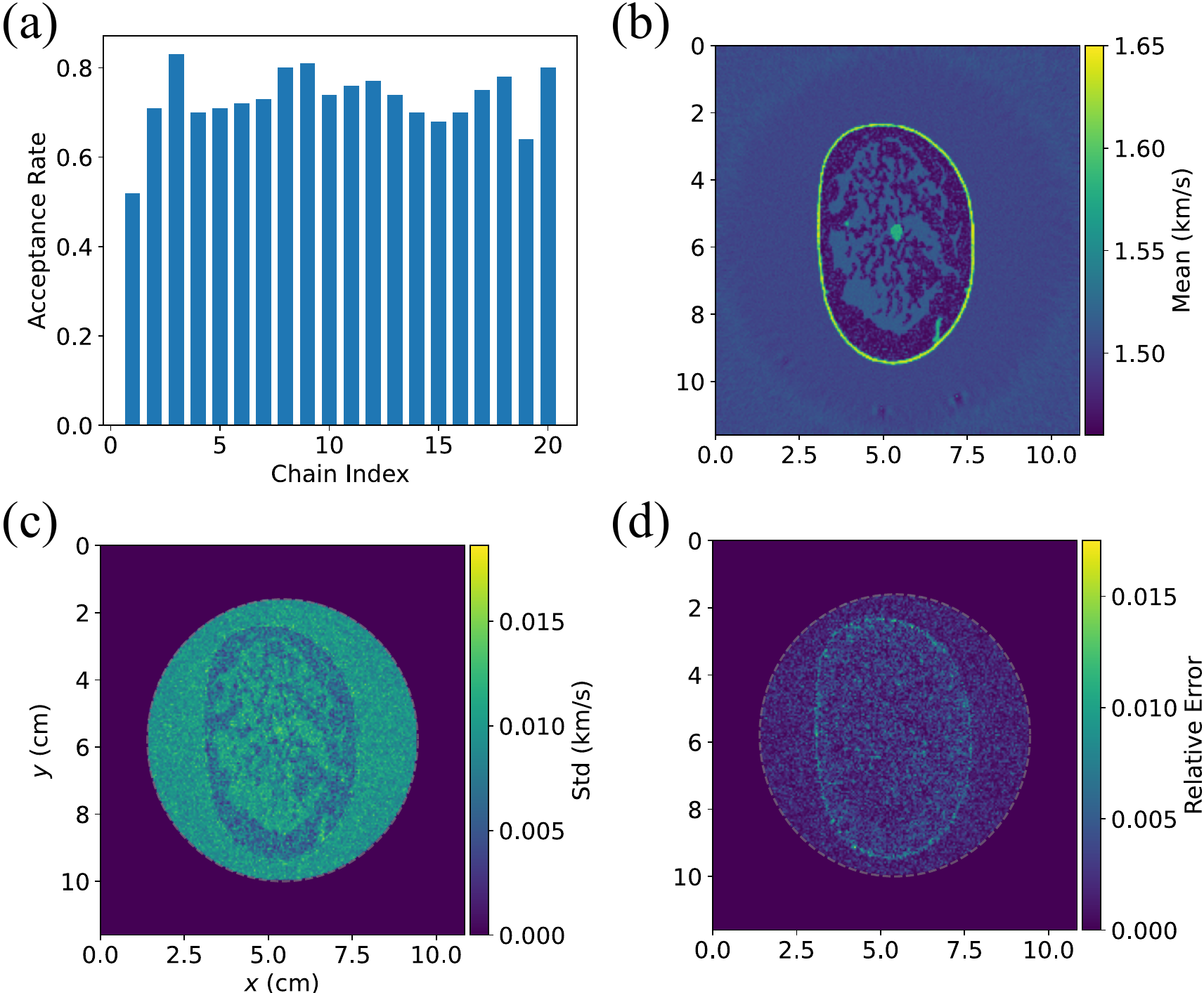}
	\caption{MH-MCMC sampling results for breast tissue SOS reconstruction. (a) Acceptance rates of 20 chains. (b) Posterior mean of the SOS obtained from 2000 samples. (c) Posterior standard deviation of the SOS. (d) Relative error of the reconstructed SOS.}
	\label{fig:fwi2DbreastMcmc}
\end{figure*}

\begin{figure*}[ht]
	\centering
	\includegraphics[width=\linewidth]{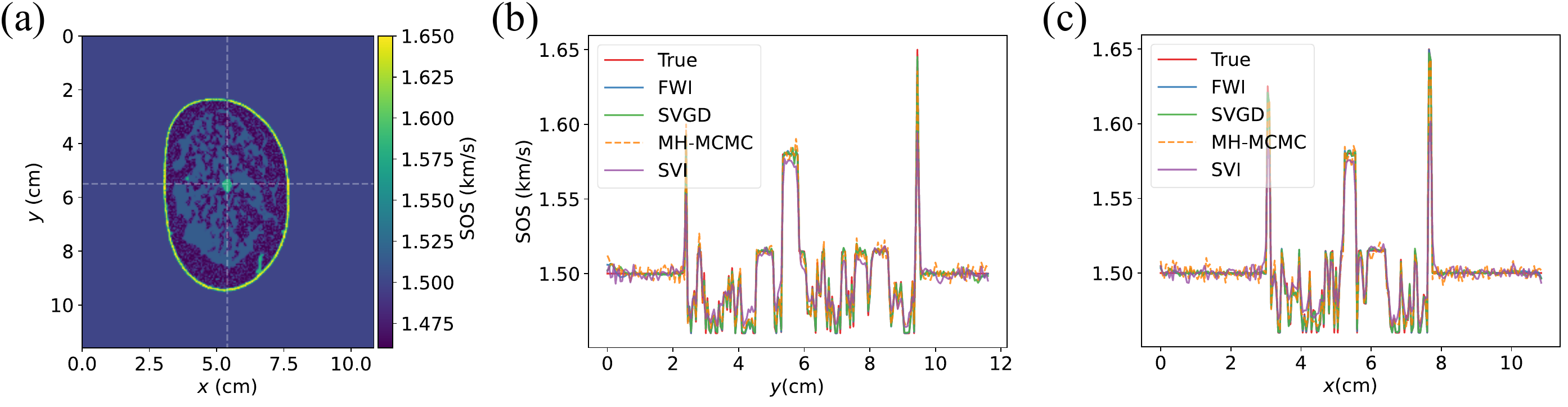}
	\caption{Variations of SOS along certain lines. 
		(a) The two selected lines at $x=5.4$ cm and $y=5.5$ cm.
		(b) and (c) show variations of SOS along the lines at $x=5.4$ cm and $y=5.5$ cm, respectively.
	}
	\label{fig:fwi2DbreastVpXYline}
\end{figure*}

After performing 100 iterations of MH-MCMC sampling, the algorithm generates 100 samples for each of the 20 chains, resulting in a total of 2000 realizations of the SOS. 
Figure~\ref{fig:fwi2DbreastMcmc}(a) shows the acceptance rates for the 20 chains, ranging from 0.52 to 0.83, with an average of 0.7295, indicating that most candidate samples are accepted.
The posterior mean and standard deviation, calculated from the 2000 samples, are presented in Figures~\ref{fig:fwi2DbreastMcmc}(b) and (c), respectively. As shown in Figure~\ref{fig:fwi2DbreastMcmc}(c), the uncertainty map within the ROI highlights the delineation of tissue boundaries. Figure~\ref{fig:fwi2DbreastMcmc}(d) presents the relative error of the reconstructed SOS, which reaches a maximum of 1.850\% and averages 0.248\% within the ROI.

The conventional FWI, SVGD, MH-MCMC, and SVI effectively reconstruct breast tissue structures. 
To evaluate local reconstruction accuracy, two cross-sectional lines at $x=5.4$ cm and $y=5.5$ cm are selected for comparison, as illustrated in Figure~\ref{fig:fwi2DbreastVpXYline}(a). 
Figure \ref{fig:fwi2DbreastVpXYline}(b) shows the reconstructed SOS for the line at $x=5.4$ cm. 
The results obtained by conventional FWI, SVGD, and MH-MCMC closely match the true value, whereas SVI yields slightly less accurate estimates in regions of higher SOS. 
Figure~\ref{fig:fwi2DbreastVpXYline}(c) presents the results along the horizontal line at $y=5.5$ cm, where SVGD again demonstrates comparable accuracy to conventional FWI.
In addition to its comparable accuracy, SVGD provides uncertainty quantification and is computationally more efficient than MH-MCMC, making it a promising method for SOS evaluation in FWI imaging.

\begin{figure*}[ht]
	\centering
	\includegraphics[width=\linewidth]{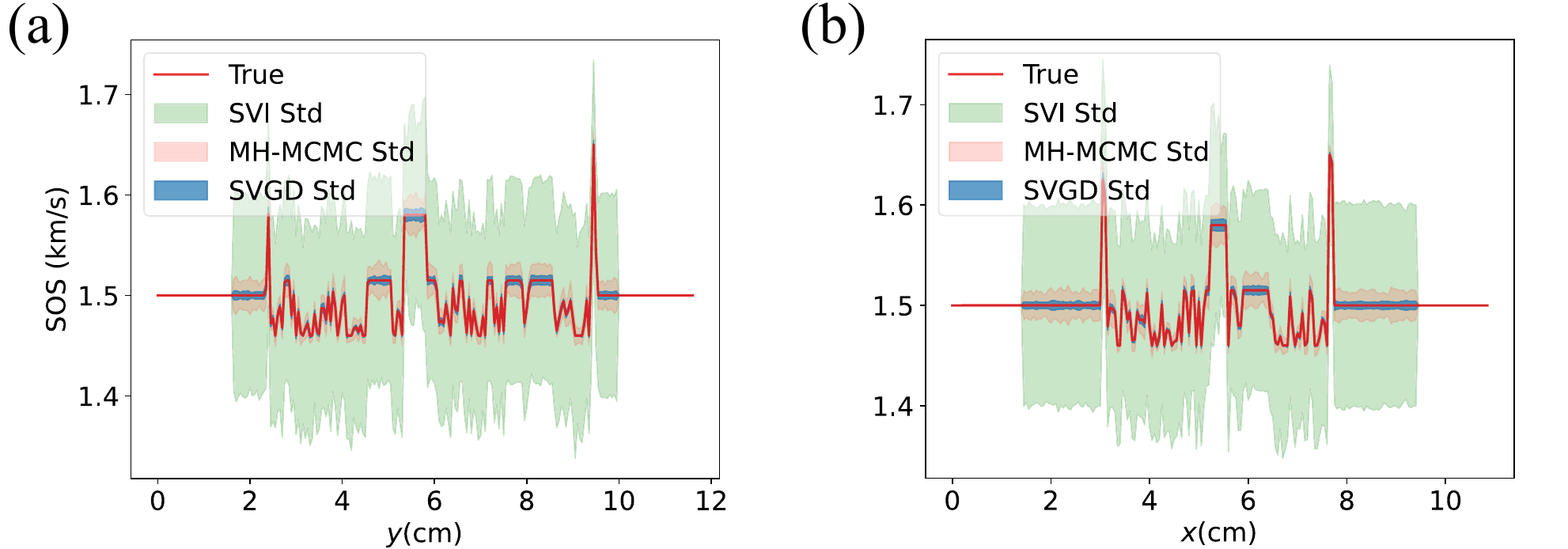}
	\caption{Variations of the standard deviation from SVGD and SVI.
		(a) and (b) show the variations of standard deviation along the lines at $x=5.4$ cm and $y=5.5$ cm, respectively.
	}
	\label{fig:fwi2DbreastStdXYline}
\end{figure*}

To evaluate the quality of uncertainty quantification produced by the two variational inference methods, SVI and SVGD, Figures~\ref{fig:fwi2DbreastStdXYline}(a) and (b) present the standard deviation profiles along the lines at $x = 5.4$ cm and $y = 5.5$ cm, respectively.
Compared to SVI, SVGD yields standard deviation estimates that better capture the variations of the true SOS, particularly in regions with higher SOS values, and exhibits strong agreement with the MH-MCMC reference.
This observation is further supported by quantitative analysis using the Pearson correlation coefficient. 
Along the line at $x = 5.4$ cm, the correlation between SVGD and MH-MCMC reaches 0.65, in contrast to only 0.01 for SVI. 
Besides, along the line at $y = 5.5$ cm, the correlation between SVGD and MH-MCMC is 0.69, whereas the correlation between SVI and MH-MCMC is 0.07.
The results demonstrate that SVGD produces uncertainty estimates more consistent with the MH-MCMC benchmark than SVI, underscoring its capability for reliable uncertainty quantification.

\subsection{Computational and memory cost analysis}

When applying the SVGD update rule Eq. (\ref{svgd update rule}) to inversion imaging, 
both computational and memory costs should be considered. The computational cost is primarily determined by the number of wave equation solves required per iteration.
In each iteration, updating all particles $\{\bm{m}_j^t\}_{j=1}^n$ requires computing the gradient of the log-posterior $\nabla_{\bm{m}_i^t} \log p\left(\bm{m}_i^t \mid \bm{d}^{\mathrm{obs}}\right)$, which consists of the likelihood gradient $\nabla_{\bm{m}} \log p\left(\bm{d}^{\mathrm{obs}} \mid \bm{m}\right)$ and the prior gradient $\nabla_{\bm{m}} \log p\left(\bm{m}\right)$. 
The likelihood gradient is computed using the adjoint-state method, requiring one forward and one adjoint wave equation solve for each emitting transducer.
In contrast, the prior gradient, derived analytically from the truncated Gaussian prior, incurs negligible computational cost. 
Consequently, each particle requires $2N_s$ wave equation solves per iteration, where $N_s$ denotes the number of emitting transducers.
For $n$ particles, the total cost per iteration is $2nN_s$ wave equation solves.
In comparison, standard deterministic FWI requires only $2N_s$ wave equation solves per iteration. 
However, SVGD enables posterior distribution approximation and uncertainty quantification, with computational cost scaling linearly with $n$.
In our implementation, we typically use $n = 20$ particles, resulting in $40N_s$ wave equation solves per iteration. Empirical results, such as those shown in Figures 5 and 10, demonstrate that using fewer particles (e.g., $n = 10$ or $15$) still yields high-quality inversion results. If moderate uncertainty resolution is sufficient, even 5 particles can provide acceptable reconstructions, thus allowing a trade-off between computational efficiency and the fidelity of uncertainty quantification.

The computational cost of the SVI-based FWI method is primarily determined by the number of wave equation solves required per iteration. 
At each iteration, $n$ samples $\{\bm{c}_i\}_{i=1}^n$ are drawn from the current Gaussian distribution. 
For each sample, the algorithm performs one forward and one adjoint wave equation solve for each emitting transducer to evaluate the misfit and compute its gradient via the adjoint-state method. 
Therefore, the total number of wave equation solves per iteration is $2nN_s$. 
Although both SVGD and SVI incur the same wave solver cost per iteration, SVGD explicitly evolves a set of interacting particles and leverages kernelized gradient updates, whereas SVI maintains and updates only the mean and covariance of the approximate posterior.

\begin{table}[!ht]
	\centering
	\caption{Comparison of computational cost over 100 iterations ($N_s=64$).}
	\label{tb: compt_cost} 
	\begin{tabular}{lcccccc}
		\toprule
		\textbf{Method} & \textbf{Particles ($n$)} & 
		\begin{tabular}[c]{@{}c@{}}Wave Solves\\ per Iteration\end{tabular} & 
		\begin{tabular}[c]{@{}c@{}}Total Wave\\ Solves (100 iters)\end{tabular} & 
		\begin{tabular}[c]{@{}c@{}}Time per\\ Iteration (s)\end{tabular} & 
		\begin{tabular}[c]{@{}c@{}}Total Time\\ (min)\end{tabular} \\
		\midrule
		FWI   & --  & $2N_s=128$      & $12{,}800$     & 26.04   & 43.4  \\
		SVGD  & 20  & $2nN_s=2560$    & $256{,}000$    & 528.6   & 881   \\
		SVI   & 20  & $2nN_s=2560$    & $256{,}000$    & 530.9   & 884.8 \\
		\bottomrule
	\end{tabular}
\end{table}

To compare the computational efficiency of different inversion methods, we apply conventional FWI, SVGD, and SVI to the breast tissue model shown in Figure~\ref{fig:fwi2DbreastSvgd}(a). 
For a fair comparison, all methods adopt the steepest descent optimization algorithm, with identical configurations for both forward and inverse modeling. 
Experiments are conducted on a high-performance computing (HPC) cluster running Ubuntu 18.04.6 LTS, equipped with four Intel Xeon E7-4830 v4 processors (112 threads in total), 503 GB of RAM, and Linux kernel version 5.4.0. 
The Devito software package (https://github.com/devitocodes/devito) is used to compute the forward wavefield and the gradient via the adjoint-state method. 
Computational times are averaged over 100 inversion iterations. As summarized in Table \ref{tb: compt_cost}, the per-iteration computational time of SVGD and SVI is approximately $n$ times that of FWI, with runtimes of 528.6~s and 530.9~s, respectively, compared to 26.04~s for FWI.
Note that the reported computational times already include the effect of $N_s=64$ emitting transducers.  

To estimate the memory requirements of the proposed SVGD-based FWI method, we analyze the dominant components per iteration for $\bm{m} \in \mathbb{R}^d$. 
First, particle storage requires maintaining $n$ particles, leading to a memory cost of $\mathcal{O}(nd)$. 
Second, the gradient of the log-posterior for each particle must be stored, also incurring a cost of $\mathcal{O}(nd)$. 
Third, computation of the kernel matrix $\mathbf{K} \in \mathbb{R}^{n \times n}$ and its gradient $\nabla_{\bm{m}_i} k(\bm{m}_i, \bm{m}_j) \in \mathbb{R}^d$ for all pairs $(i, j)$ results in a memory cost of $\mathcal{O}(n^2d)$. 
Fourth, each particle must store forward and adjoint wavefields across a space-time grid. The dominant term for each wavefield is $\mathcal{O}(dT)$, yielding a total memory cost of $\mathcal{O}(nN_r dT)$ across all particles, where $N_r$ is the number of receiving transducers.
Since the method employs single-source emission, the number of emitting transducers $N_s$ does not significantly affect memory storage for the wavefields.
Other temporary variables, such as misfit values or prior terms, contribute negligibly to the overall cost. 
Summing the dominant terms, the total memory requirement is $\mathcal{O}(nd(1 + n + T + N_r))$, which scales linearly with $d$, $T$, and $N_r$, and quadratically with $n$.
For moderate values of $n$ (e.g., 10–20), the method remains memory-efficient relative to conventional ensemble-based Bayesian sampling approaches.

\section{Discussion} \label{sec: discussion}

FWI is a powerful technique for imaging  biological tissue properties, yet estimating uncertainty and solving Bayes' equation for high-dimensional acoustic data is computationally expensive. 
This study tackles this challenge by using the SVGD algorithm. 
By combining the adjoint method with a Gaussian prior assumption, Bayes' equation is solved to obtain the posterior distribution of the parameters. 
The efficiency of the SVGD algorithm in FWI depends on the choice of prior distribution.
Selecting an appropriate prior can accelerate the algorithm's ability to produce accurate results. 
One possible solution for acceleration is to incorporate deep learning models, such as Variational Autoencoders (VAE) \cite{liang2024survey,zhou2024transvae,vahdat2020nvae} or Generative Adversarial Networks (GANs) \cite{rocha2024vae,ali2024generative,foo2024tumor}, which learns the parameter distribution from medical imaging data like MRI or USCT. 
These models generate more realistic priors, further improving the performance of the SVGD algorithm in complex inversion tasks.

A practical limitation of the SVGD-based FWI framework lies in its computational cost, particularly for large-scale inverse problems. Each particle requires one forward and one adjoint wave equation solve per iteration, resulting in a total of $2n$ wave equation solves. Although this increases the computational burden compared to deterministic FWI, empirical results (e.g., Figures \ref{fig:fwi2DringSvgdNum_1} and \ref{fig:fwi2DbreastSvgdNum_1}) demonstrate that satisfactory inversion accuracy and uncertainty quantification can be achieved with as few as 5–10 particles. This flexibility offers a trade-off between computational efficiency and the fidelity of the posterior distribution. Additionally, since the forward and adjoint wave equation solves for different particles are independent, the method is inherently parallelizable. By leveraging high-performance computing resources and executing computations in parallel across particles, the wall-clock time can be significantly reduced, making SVGD-based inversion feasible for more complex and computationally demanding imaging tasks.

The SVGD algorithm offers the advantage of quantifying uncertainty in the imaged regions, allowing for better decision-making when interpreting the results. 
A key observation is that the lower standard deviation, the more reliable the inversion result for that region. Areas with higher standard deviation suggest that the solution may be unstable or unclear. 
The high uncertainty could be mitigated by giving more importance to regions with lower standard deviation. One potential strategy is to introduce an adaptive weighting scheme that assigns lower weights to noisy data and higher weights to cleaner data. This scheme is further integrated into the loss function to minimize the impact of noisy data. 
Besides, incorporating techniques such as stochastic gradient descent (SGD) \cite{azimjonov2024stochastic,beznosikov2023stochastic,sclocchi2024different} or mini-batch gradient estimation \cite{qi2023statistical,singh2024mini,wang2024analysis}, the adaptive weighting scheme ensures that more important subsets of the data are used in each iteration, helping to stabilize the particle updates. 
In regions of high uncertainty, applying additional methods such as smoothing, noise filtering, or regularization helps refine the parameter estimates, ultimately improving the overall resolution of the FWI.

In clinical diagnosis, the standard deviation may accelerate artificial intelligence (AI) diagnosis by highlighting uncertain results caused by noise, artifacts, or insufficient data. The ability to quantify uncertainty allows AI systems to flag regions that need further examination or additional data collection. By focusing on regions with high uncertainty, AI guides clinicians in prioritizing additional imaging or confirming diagnoses. This capability helps reduce the likelihood of misdiagnosis and supports more informed clinical decision-making. By providing a reliable measure of uncertainty, the standard deviation ultimately enhances the trustworthiness of FWI-based imaging techniques, making them more effective in real-world clinical applications.

\section{Conclusion} \label{sec: conclusion}

In this study, an SVGD-based FWI algorithm is employed for medical ultrasound imaging, demonstrating its effectiveness in reconstructing the SOS distributions of imaged objects.
For comparison, the conventional FWI and SVI are also used to reconstruct the SOS.
The results show that all three algorithms provide accurate estimates of object properties, with SVGD offering additional benefits, such as faster convergence and the ability to quantify uncertainty. 
The SVGD results exhibit low relative errors and reliable uncertainty in the core regions of the object. 
Furthermore, the performance of three algorithms is evaluated using a realistic breast tissue model.
The SVGD algorithm demonstrates comparable or even superior accuracy to other two algorithms.
Moreover, the uncertainty quantified by SVGD aligns closely with the benchmark MH-MCMC method, reflecting SOS variations of the inversion imaging. Sensitivity analysis of particle count shows that even with fewer particles (e.g., $n = 10$ or $15$), high-quality inversion results are achievable. When moderate uncertainty resolution is acceptable, 5 particles can provide satisfactory reconstructions, thereby enabling a balance between computational efficiency and the fidelity of uncertainty quantification. The results underscore the potential of SVGD-based FWI as a promising tool for inversion imaging tasks.

Future work focus on incorporating prior information into the SVGD algorithm. 
By adding covariance to the prior, the accuracy and efficiency of the reconstruction process may be further improved.
Additionally, integrating shear wave data with the SVGD algorithm could enhance elasticity imaging, particularly for assessing tissue stiffness, which is critical in evaluating conditions such as musculoskeletal tissue disorders and tumors. 
Combining SVGD with shear wave data may also help mitigate the effects of cycle-skipping near bone tissue, a challenge often encountered in FWI. 
This approach may require fine-tuning and adaptation of the algorithm to address specific challenges in different scenarios, but it holds potential for enhancing the diagnostic capabilities of ultrasound and other imaging modalities.

\section{Acknowledgements}
This work was supported by the National Natural Science Foundation of China (grant numbers 12122403, 12034005, and 12327807) and National Key R\&D Program of China (2023YFC2410800).
The computations for this research were performed using the CFFF platform at Fudan University.
Thanks to Oscar Bates for his constructive suggestions.

\section{Declaration of Competing Interest}
The authors declare that there are no known competing interests that could have influenced the work reported in this paper.

\bibliographystyle{unsrt}
\bibliography{bibliography}

\end{document}